\documentclass[fleqn,usenatbib]{mnras}
\usepackage[T1]{fontenc}
\usepackage{ae,aecompl}
\usepackage{graphicx}
\usepackage{amsmath}
\usepackage{amssymb}
\usepackage{txfonts}

\title[Resequencing the Hubble sequence]{Resequencing the Hubble sequence 
and the quadratic (black hole mass)-(spheroid stellar mass) relation for
elliptical galaxies}

\author[Graham]
{
Alister W.\ Graham$^{1}$\thanks{E-mail: AGraham@swin.edu.au}
$^1$ OzGrav-Swinburne, Centre for Astrophysics and Supercomputing, Swinburne
University of Technology, Hawthorn, VIC 3122, Australia
}

\date{Accepted XXX. Received YYY; in original form ZZZ}
\pubyear{2022}

\begin{document}
\label{firstpage}
\pagerange{\pageref{firstpage}--\pageref{lastpage}}
\maketitle

\begin{abstract}
One of the most protracted problems in astronomy has been 
understanding the evolution of galaxy morphology. 
Much discussion has surrounded how lenticular galaxies may form a bridging
population between elliptical and spiral galaxies.  However, with recourse to
a galaxy's central black hole mass, accretion-built spiral galaxies have
emerged as the bridging population between low-mass lenticular galaxies and
the dusty merger-built lenticular galaxies contiguous with elliptical galaxies
and `brightest cluster galaxies' in the black hole/galaxy mass diagram.  
Spiral galaxies, including the Milky Way, 
appear built from gas accretion and minor mergers onto what
were initially lenticular galaxies.  These connections are expressed as a new
morphology sequence, dubbed the `Triangal', which subsumes elements of the
Hubble sequence and the van den Bergh trident and reveals the bridging nature
of the often overlooked ellicular galaxies.  Furthermore, a quadratic
black hole/galaxy mass relation is found to describe ordinary elliptical
galaxies.  The relation is roughly parallel to the quadratic-like relations
observed for the central spheroidal component of spiral galaxies, dust-rich
lenticular galaxies, and old dust-poor lenticular galaxies.  The brightest
cluster galaxies are offset according to expectations from an additional major
merger.  The findings have implications for feedback from active galactic
nuclei, mapping morphology into simulations, and predicting gravitational wave
signals from colliding supermassive black holes.  A new galaxy speciation
model is presented. It disfavours the `monolithic collapse' scenario for
spiral, dusty lenticular, and elliptical galaxies.  It reveals substantial
orbital angular momentum in the Universe's first galaxies and unites dwarf and
ordinary `early-type' galaxies.
\end{abstract}


\section{Introduction}
\label{Sec_Intro}

Since the first spiral `nebula' was discovered \citep{1850RSPT..140..499R},
astronomers have pondered their formation and connection with other 
nebulae \citep{1852AJ......2...95A, 1895MNRAS..56...70R, 
1906PASP...18..111A, 1919pcsd.book.....J}.  
Indeed, how the extragalactic nebulae, the  `island universes'\footnote{\citet{1845kosm.book.....V} 
used the German word Weltinsel (world island) to refer to our Galaxy and
everything floating in space.  \citet{1846AN.....24..213M} subsequently referred
  to the nebulae as ``world islands'',  
which \citet{Mitchel1847} translated and adapted into  ``island universes''.}
of \citet{1917PASP...29..206C}, 
nowadays referred to as galaxies, may evolve and transform 
from their primordial incarnation remains an active and open topic of research
\citep[e.g.,][]{2006PASP..118..517B, 2008ApJ...674..784P, 2008MNRAS.387...79V,
  2012MNRAS.425...44C, 2014ARA&A..52..291C, 2014MNRAS.440..889S}. 
With roots in the 1700s, today's most well-known
galaxy sequence was stripped of its evolutionary pathways almost
as soon as it was conceived a century ago. 
As revealed in the well-referenced papers by 
\citet{1971JHA.....2..109H} and \citet{2013ASPC..471...97W}, see
also \citet{2019MNRAS.487.4995G}, credit for the E-to-S 
\citep{1926ApJ....64..321H} and E-S0-S \citep{1936rene.book.....H}
`Hubble sequence' resides with many. 

\citet{1919pcsd.book.....J} popularised the notion of amorphous 
round/elliptical-shaped nebulae evolving into ringed/spiral
nebulae.  Such a concept originated from 
the 18th century `nebular hypothesis' in which 
elliptical-shaped nebulae were thought to 
rotate and throw-off rings and spiral arms at the expense of a dwindling
central bulge. 
Building on this, 
\citet{1920MNRAS..80..746R} introduced the essence of the 
early-to-late type spiral sequence, 
based in part\footnote{An additional facet was the appearance of
  condensations in the arms.  This facet stemmed from earlier speculation that planets
  may condense out of material ejected from nebulae, and thus a more
  granulated appearance reflected a more evolved system. As
  \citet{1971JHA.....2..109H} note, in 1923, Hubble embraced this idea, along
  with noting the 
  presence/absence of a spiral galaxy's bar, 
  as previously reported by \citet{1918PLicO..13....9C}.}  
on the dominance of the bulge, 
with this central concentration later adopted as criteria by 
\citet{1925MNRAS..85..865L, 1926ArMAF..19B...8L} and \citet{1926ApJ....64..321H}. 
The notion of an initial (early) 
and latter (late) type of nebulae was initially entertained by
Hubble\footnote{As noted by \citet{1971JHA.....2..109H}, in 1923, in preparation for the second 
  International Astronomical Union, Hubble submitted a manuscript to
  Slipher in which he
  wrote ``there is some justification in considering the elliptical nebulae as
representing an earlier stage of evolution'' and he also now listed them {\it before} the
spiral nebula, reversing the spiral-spindle/ovate order in his original scheme
\citep{1922ApJ....56..162H}.} 
but later disfavoured by many in the early-to-mid 1920s, 
including \citet{1926ApJ....64..321H}.  Nonetheless, the early- and late-type
nomenclature was retained.  
The practice of quantifying the roundness of a `regular' nebula --- encompassing
the early-type 
galaxies --- with designations ranging from 1 for round nebulae to 5 for
elongated nebulae had been used since \citet{1847raom.book.....H}. 
While Hubble expanded upon this practice, 
he also sought to distil the key elements from the detailed scheme of  
\citet{1908PAIKH...3..109W}.\footnote{While the nebulae classification scheme
  of \citet{1908PAIKH...3..109W} is not in use, Charles Wolf is 
  remembered through Wolf-Rayet stars \citep{1867CRAS...65..292W}, 
  the central stars of {\it planetary} nebulae \citep{1914ApJ....40..466W}.}
Furthermore, \citet{1918PLicO..13....9C} had recently introduced the 
barred versus non-barred designation 
following the murky identification of bars by \citet{1915HelOB..15..129K}. 
This led to the bifurcation of the S galaxies that 
\citet{1926ApJ....64..321H} included in his first table. 
\citet[][their footnote~42]{1971JHA.....2..109H} and 
\citet{1997AJ....113.2054V} remind us that 
\citet{1928asco.book.....J} was the first to express this visually, as a Y-shaped diagram before
\citet{1936rene.book.....H} turned it sideways to give us the so-called
`tuning fork', with the intermediary spindle/lenticular class, i.e., the 
armless disc galaxies 
introduced by \citet{1925MNRAS..85.1014R}, being the S0 galaxies at the junction.

The longevity of the Jeans/Reynolds/Hubble sequence 
of galaxies, encapsulated by the tuning fork 
 --- taught in most 
introductory astronomy courses --- is perhaps surprising given that doubt over
the direction of potential morphological transformations led \citet{1926ApJ....64..321H} to recant
that this was an evolutionary sequence \citep{1971JHA.....2..109H}.  However, 
a wealth of additional characteristics was subsequently grafted onto
and coded in by 
\citet{1959HDP....53..275D} and \citet{2007dvag.book.....B}, and this does 
reflect some of the formation histories of galaxies \citep{2013pss6.book....1B}. 

While some galaxies, and their spheroidal component, are known to be 
built by collisions \citep[e.g.,][]{2003ApJ...597..893N, 2006ApJ...648..976M,
  2022ApJ...940..168C}, vital clues from the coevolution of their massive
black holes (BHs) have only now come to light.  It has recently been revealed
how gas-poor, aka `dry', mergers of S0 galaxies \citep{1925MNRAS..85.1014R,
  2009ApJ...702.1502V} have created the offset population of E galaxies
(including brightest cluster galaxies, BCGs) in the diagram of BH mass,
$M_{\rm bh}$, versus spheroid\footnote{Here, the spheroidal component of a
  galaxy refers to either a central bulge of a disc galaxy or the bulk of an E
  galaxy.}  stellar mass, $M_{\rm *.sph}$ \citep{Graham:Sahu:22a}.  The
merger-induced transition of stars from ordered rotating discs in S0 galaxies
into a somewhat chaotic `dynamically hot' spheroidal-shaped swarm, coupled
with the steep $M_{\rm bh}$-$M_{\rm *.sph}$ mass scaling relation for S0
galaxies, explains why, at a given mass, the E galaxies have an
order-of-magnitude lower $M_{\rm bh}/M_{\rm *.sph}$ ratio than the S0
galaxies, as first observed by \citet{2019ApJ...876..155S}.

Feedback from `active galactic nuclei' 
\citep[e.g.,][]{1964ApJ...140..796S, 1998A&A...331L...1S, 2014ARA&A..52..589H} 
has typically been heralded 
as the driving force behind the BH/galaxy scaling relations 
\citep[e.g.,][]{1998AJ....115.2285M, 2000ApJ...539L...9F, 2000ApJ...539L..13G, 2001ApJ...563L..11G},
with the initially small scatter about the $M_{\rm bh}$-(stellar velocity
dispersion, $\sigma$) relation 
taken as proof.  However, it is now apparent that dry mergers, rather than BH 
feedback, have dictated the behaviour of the E galaxies in the $M_{\rm
  bh}$-$M_{\rm *,sph}$ diagram \citep{Graham:Sahu:22a}.  Furthermore, the virial
theorem, coupled with the best measurements of spheroid size and stellar mass,
has revealed how dry mergers explain why the $M_{\rm bh}$-$\sigma$ relation
does not have much scatter at the high-mass end where the E galaxies reside
\citep{Graham-sigma}.

Inspecting {\it Hubble Space Telescope} ({\it HST}) images available at the Hubble Legacy
Archive (HLA)\footnote{\url{https://hla.stsci.edu/}}, it has also recently
been revealed that there are two populations of S0 galaxy: dust-poor and
dust-rich \citep{Graham-S0}.  Based on the results in Appendix~\ref{Apdx1},
this may mirror a division previously detected as 
low- and high-luminosity S0 galaxies \citep{1990ApJ...348...57V}, which 
needed to be explained and integrated into a joint evolutionary and 
galaxy morphology classification scheme.  The existence of two 
populations helps explain a century of confusion, different formation
scenarios, and physical properties for S0 galaxies 
\citep{2012AdAst2012E..28A}.  The dusty S0 galaxies are major-merger remnants
that involved gas and star formation, referred to as `wet' mergers.  As the
star formation fades, these S0 galaxies will migrate across the `green valley'
and on to the `red sequence' in diagrams of colour versus stellar mass
\citep{2017ApJ...835...22P}.  The S galaxies are observed to reside between
the dust-poor and dust-rich S0 galaxies in the $M_{\rm bh}$-$M_{\rm *.sph}$
diagram.  Furthermore, the S galaxies are known to have been built, or rather renovated, by
minor mergers, which may encompass the accretion of gas clouds from
surrounding H{\footnotesize I} \citep[e.g.,][]{2007IAUS..235...29B,
  2010ApJ...725.1550C, 2015MNRAS.453.2399W}, the devouring of satellite
galaxies, and the capture of dwarf galaxies 
\citep[e.g.,][]{1978ApJ...225..357S, 1997AJ....114.1858P, 2010ASPC..423....3G, 2018ApJ...852...44G,
  2018ApJ...866...22L, 2019MNRAS.486.3180K, 2021ApJ...907...85M}.
While such gravitational disturbances may invoke a spiral pattern 
\citep{1966ApJ...146..810J}, too large a merger is likely to dynamically
overheat the disc, destroy or prevent any spiral,
and produce a dusty S0 galaxy such as NGC~3108 \citep{2008MNRAS.385.1965H} or Centaurus~A 
\citep{1983PASP...95..675E}, which, in time, may resemble something more like
the Sombrero galaxy.

\begin{figure*}
\begin{center}
\includegraphics[trim=0.0cm 0cm 0.0cm 0cm, width=0.66\textwidth, angle=0]{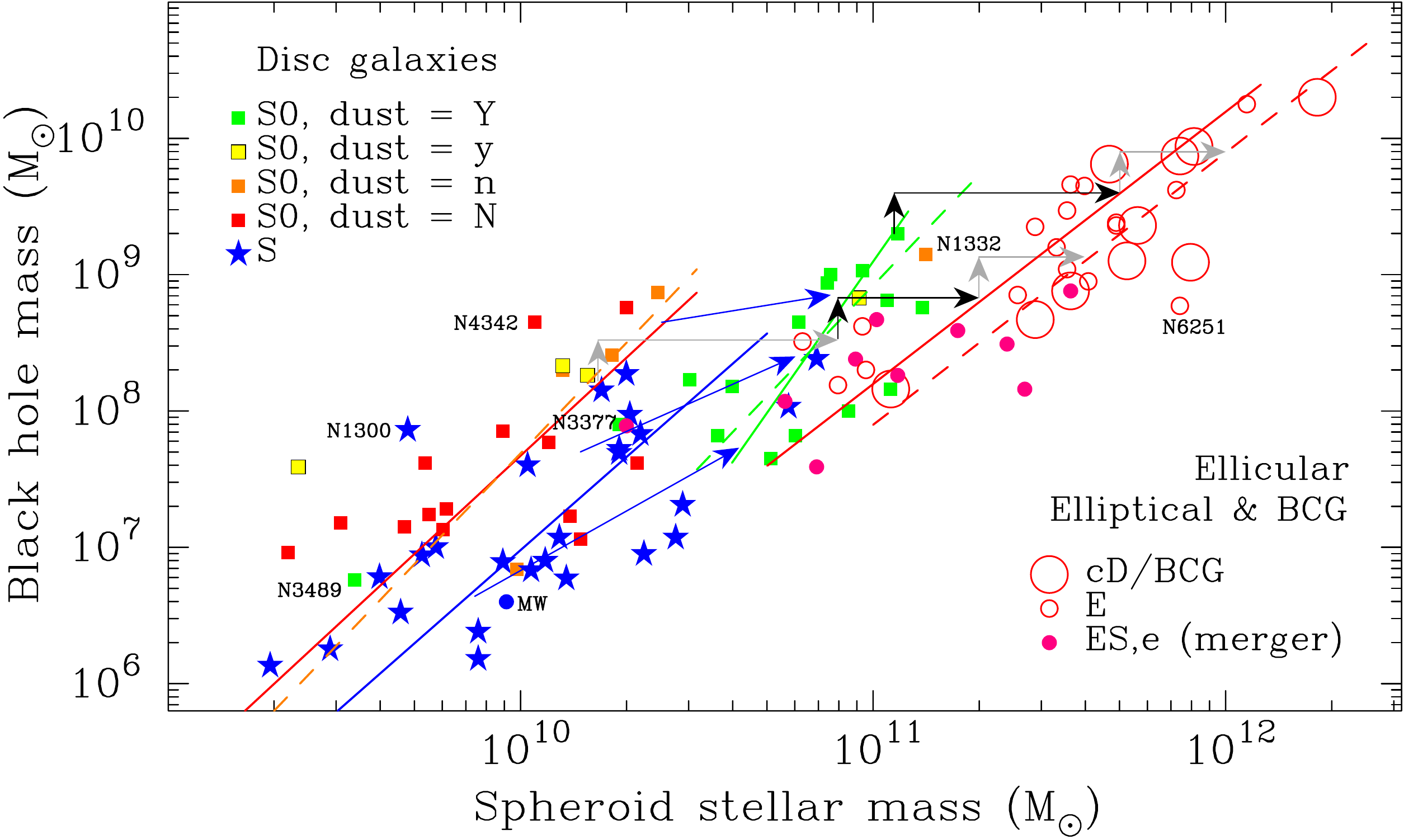}
\caption{Morphologically-aware $M_{\rm bh}$-$M_{\rm *,sph}$ diagram.
The ten cD and BCG (including the dusty S0 galaxy NGC~1316 
and the ES,e galaxy NGC~1275), along with NGC~3377 and NGC~6251, were excluded from the fit to 
the (remaining) 24 E/ES,e galaxies (right-most solid red line: Eq.~\ref{Eq1}). 
The one BCG above this line is NGC~4486. 
The (non-BCG) E galaxy with the highest BH mass is NGC~1600.
The dashed red line represents the BCG. 
The lines for the (S0 and S) disc galaxies have come from \citep{Graham-S0}.
Using the S0 galaxy `dust bins' \citep{Graham-S0}, 
the left-most red solid line represents S0 galaxies without visible signs of
dust (dust = N),
while the orange dashed line additionally includes S0 galaxies with only 
a nuclear dust disc or ring (dust = n). 
The green dashed line is the orange dashed line 
shifted horizontally by an arbitrary $\log(3.5)\approx0.54$ dex, while the
solid green 
line is a fit to the dusty (dust = Y) S0 galaxies, excluding those with
only a little widespread dust (dust = y). 
The blue line represents the S galaxy data. 
Labelled galaxies were excluded from the Bayesian analyses. 
From left to right, the logarithmic slopes are: 
2.39$\pm$0.81 (red line); 
2.70$\pm$0.77 (dashed orange line); 
2.27$\pm$0.48 (blue line); 
3.69$\pm$1.51 (solid green line); 
2.70$\pm$0.77 (dashed green line); and 
2.00$\pm$0.25 (red solid and dashed line). 
Full equations are in Table~\ref{Table1}, and the arrows are explained in the main text.}
\label{Fig1}
\end{center}
\end{figure*}

Given that the BCGs are likely to be E galaxies built by multiple mergers 
\citep[e.g.,][]{2003AJ....125..478L}, 
the E galaxy population is explored more thoroughly here.  The BCGs may be offset
in the $M_{\rm bh}$-$M_{\rm *.sph}$ diagram from the (non-BCG, or simply
`ordinary') E galaxies.  An offset will not be observed if the E galaxies follow
the near-linear $M_{\rm bh}$-$M_{\rm *.sph}$ relation they have been thought
to follow for a quarter of a century 
\citep{1998AJ....115.2285M,2013ARA&A..51..511K,2016ApJ...818...47S}.  However,
for a steeper than linear non-BCG E galaxy $M_{\rm bh}$-$M_{\rm *.sph}$ relation, 
the addition of two ordinary E galaxies' stars and their central BH
would lead to a merger-induced jump, referred to as `punctuated equilibrium'
\citep{Graham:Sahu:22a}\footnote{As \citet{Graham:Sahu:22a} noted, 
the phrase `punctuated equilibrium' stems from studies of Darwinian evolution.
\citet{Eldredge_Gould_1972} proposed that species
are stable until a rare and rapid event occurs.  
For a galaxy, such an event could be represented by a 
substantial merger, resulting in significant evolutionary change. 
Within evolutionary biology, such branching speciation is called
cladogenesis.}, 
taking them off the ordinary E galaxy  $M_{\rm bh}$-$M_{\rm *,sph}$
relation.  This jump is detected here, and, for the first time, all of the merger-built morphological
transformations are shown to map into a triangular-like diagram revealing fundamental
connections between the galaxy types.  The `Triangal', presented herein, supersedes the Hubble
sequence by (i) redrawing the connections and (ii) including evolutionary pathways. It
is, nonetheless, a development based on the works of many, in particular 
\citet{1976ApJ...206..883V, 1990ApJ...348...57V}.

\section{Data: Ordinary elliptical galaxies versus BCGs}

The data for this investigation consists of published 
BH masses, spheroid (and galaxy) stellar masses, and the
galaxies' morphological type, including whether the E galaxies are BCG or
cD\footnote{cD galaxies may be the first or second brightest galaxy in a
  cluster.  They have a centrally-dominant location, and as such, they are
  immersed at the centre of the intracluster light, which appears as a diffuse
  halo \citep{2007ApJ...668..826C}.} \citep{Graham:Sahu:22a, Graham:Sahu:22b},
and S0 galaxy `dust bin' \citep{Graham-S0}. 
For ease of reference, the cD and BCG will often collectively be
referred to as BCG in the text. 
The spheroid masses were obtained from careful\footnote{Rather than blindly
  fitting multiple S\'ersic functions, (galaxy component)-specific functions
  were fit after inspecting the images and consulting with the literature,
  including kinematic maps.} multicomponent decompositions,
which separate bars and inner discs from spheroids and other galaxy
components.  X/(peanut shell)-shaped structures were captured by a Fourier
analysis of the isophotal shapes 
\citep{1978MNRAS.182..797C, 2015ApJ...810..120C, 2016MNRAS.459.1276C} 
and effectively folded back into the bar component.  As such, components that
some may call a `pseudobulge' or a false bulge are not considered 
the spheroid.  The spheroid may, however, have a S\'ersic index of less than
2$\pm$0.5, a divide that has been questioned \citep[e.g.,][]{2013pss6.book...91G,
  2019PASA...36...35G}. 
For each galaxy, the decomposition has been plotted and published 
\citep{2016ApJS..222...10S,2019ApJ...876..155S,2019ApJ...873...85D,Graham:Sahu:22b}.

\section{Results}

The current focus explores separating the spheroid-dominated E and ES,e
galaxies\footnote{Ellicular (ES) galaxies \citep{1966ApJ...146...28L} are
  spheroid-dominated but have fully embedded discs.  Also known as `disc
  ellipticals' \citep{1988A&A...195L...1N}, \citet[][their
    table~1]{2015ApJS..217...32B} referred to them as S0$^-$ sp/E5-E7 and 
  presented a dozen examples in their figure~23.  They typically have
  spheroid-to-total stellar mass ratios around $\sim$0.9$\pm0.05$
  \citep{Graham:Sahu:22a}, and there might be two species 
  \citep{Graham:Sahu:22b}.  The ES,b galaxies have spheroids more akin to the
  bulges of S0 galaxies, and the four in the current sample are treated as
  such here.  They may be relics from the early merging of the now-dust-poor
  S0 galaxies with high bulge-to-total stellar mass ratios ($B/T$), or old
  spheroid-dominated systems which only accreted an intermediate-scale disc.
  The ES,e galaxies have spheroids more like E galaxies, and are likely built from the
  merger of dusty S0 and/or gas-poor S galaxies.}  into BCGs and non-BCGs
\citep{Graham:Sahu:22b}.  Given that galaxy groups can be small in number (3-5), a
brightest group galaxy 
(BGG) may be an ordinary E galaxy, a dusty S0 galaxy or occasionally an S galaxy
if no major merging has occurred. As such, the BGG tend not to not distinguish
themselves in the $M_{\rm bh}$-$M_{\rm *.sph}$ diagram.
As indicated in Section~\ref{Sec_Intro}, 
the author has been evolving this diagram piecewise to make the changes more
digestible and to better emphasise the importance of galaxy morphology and
origin. 
Readers familiar with only a single regression line in this diagram may like
to review figure~1 in \citet{Graham:Sahu:22b}, which separates the galaxies into
S, S0 and E types, and figure~4 in \citet{Graham-S0}, which further separates
the S0 galaxies into dust-poor and dust-rich bins.  These `dust bins' or
classes are illustrated in figure~1 of \citet{Graham-S0}.\footnote{A more
  quantitative approach could involve the specific dust mass
  \citep[e.g.][]{2008ApJ...678..804E}, 
 i.e. the dust mass normalised by 
  the stellar mass, but one may still need to pay attention to nuclear dust
  discs versus wide-spread dust.} 
In what follows, the
BCGs (most of which are E galaxies) are separated from the ordinary (non-BCG) E
galaxies. 

Figure~\ref{Fig1} shows the ordinary E galaxies \citep[including the ES,e
  galaxies described and identified in][]{Graham:Sahu:22b} and ten BCGs in the
 $M_{\rm bh}$-$M_{\rm *,sph}$ diagram.
Due to
their ability to skew the result, two apparent outliers --- NGC~3377 (ES,e)
and NGC~6251 (E), which are marked in Figure~\ref{Fig1} and discussed in 
Appendix~\ref{Apdx2} along with other interesting outliers --- are
excluded from the Bayesian analysis \citep[method described
  in][]{2019ApJ...873...85D} performed here on the 
ordinary E and ES,e galaxies. This analysis yields\footnote{Including NGC~6251
  gives a logarithmic slope of 1.90$\pm$0.25, while including NGC~3377 and NGC~6251 gives
  a logarithmic slope of 1.65$\pm$0.22.}
\begin{equation}
\log(M_{\rm bh}/M_\odot)= (2.00\pm0.25)[\log(M_{\rm *,sph}/M_\odot)-11.32]+(8.84\pm0.15). 
\label{Eq1}
\end{equation}
This quadratic relation is dramatically different to the near-linear relation 
previously thought to define the coevolution of
E galaxies and their BHs
\citep{1998AJ....115.2285M}. 

The bulk of the ten BCGs are seen to reside to the right of Eq.~\ref{Eq1} in Figure~\ref{Fig1}. 
This is readily explained if, on
average, they predominantly formed from the major merger of two E or ES,e
galaxies.  The situation could equally represent the merger of several S0 galaxies or some
other suitable combination.  The steeper than linear nature of the ordinary
E/ES,e $M_{\rm bh}$-$M_{\rm *,sph}$ relation (Eq.~\ref{Eq1}) results in dry mergers
forming BCGs that reside to the right of, rather than on, Eq.~\ref{Eq1}.  More 
data are required to obtain a reliable fit for the distribution of the BCGs.
Therefore, a simple mass doubling from a major dry merger has been used to
define the dashed red line in Figure~\ref{Fig1}, which appears broadly representative of
the distribution of the BCGs.

Differing from the single S0 galaxy $M_{\rm bh}$-$M_{\rm *,sph}$
relation presented in \citet{Graham:Sahu:22a}, the S0 galaxies are placed
in four `dust bins' following \citet{Graham-S0}. These bins are denoted as
follows: 
N for no visible dust;
n for only a nuclear dust disc/ring;
y for weak widespread dust; and
Y for strong, widespread dust features. 
As alluded to in the Introduction, 
this was established in \citet{Graham-S0} by looking at colour images in the HLA. 
Roughly, nuclear discs are less than a few hundred parsecs in (radial) size, while
widespread features may be 2 to 3 or more kpc in (radial) size.
To a certain degree, one can expect a general sequence of increasing dust from the
low-mass S0 galaxies to the spiral galaxies and on to the (wet merger)-built
S0 galaxies, 
some of which were previously ultraluminous infrared galaxies
\citep[ULIRGs:][]{2003ApJ...582L..15K, 2006ApJ...638..745D}. 
This stems from the trickle of star formation in spiral galaxies and the
starbursts which formed the dusty S0 galaxies, in which metals condensed out
of the interstellar medium. 

The dust-poor S0 galaxies need not be gas-poor, and some may contain expansive
H{\footnotesize I} envelopes \citep[e.g.,][]{1995AJ....109..990V}. 
Massive reservoirs of hydrogen gas are known to surround some low-mass and low
surface brightness galaxies \citep[e.g.,][]{1993AJ....106...39H,
  1997ARA&A..35..267I, 2000ApJ...541..675B}.  Low surface brightness 
galaxies are also known to be metal-poor \citep[e.g.,][]{1994ApJ...426..135M}. 
Such gas clouds may remain 
indefinitely unless an angular-momentum-robbing gravitational disturbance
drives them inward to fuel a galactic metamorphosis or a passing neighbour
leads to a gas bridge. These gas clouds could instead be removed via several
well-known mechanisms within a group or cluster environment. The latter
processes will leave a dust-poor S0 galaxy, while the first may build an S
galaxy. 
In passing, it is noted that a dust-poor S0 galaxies' present-day mass function of stars
will be a truncated and modified form of the initial mass function from 10-13
Gyrs ago. With stellar mass loss due to winds and supernovae ejecta, coupled
with ram-pressure stripping within a group/cluster environment, 
the galaxies' stellar masses will reduce over time. 
Thus, the stellar orbits within the 
galaxies' discs will slightly expand from their initial configuration. Coupled
with a faded stellar population, the surface brightnesses will be reduced.  
That is, this `bloating' (in the plane) 
of the disc --- after consumption of the available gas at the formation epoch ---
adds to the dimness of 
local ($z \sim 0$) low surface brightness and ultra-diffuse disc galaxies (and dwarf
spheroidal-shaped galaxies).

In Figure~\ref{Fig1} , relations for the S0 galaxies with either no dust or
strong dust features are included for reference with the relation for non-BCG
E galaxies. Equations are provided in Table~\ref{Table1}.

\begin{table}
\centering
\caption{$M_{\rm bh}$-$M_{\rm *,sph}$ and $M_{\rm bh}$-$M_{\rm *,gal}$ relations}
\label{Table1}
\begin{tabular}{lrcccc}
\hline
Galaxy type  &  $N$ &  slope (A) & mid-pt (C) & intercept (B) & $\Delta_{\rm rms}$ \\
\hline
\multicolumn{6}{c}{$\log(M_{\rm bh}/{\rm M}_\odot) = A[\log(M_{\rm
      *,sph}/\upsilon\,{\rm M}_\odot) - C] +B$ (Figure~\ref{Fig1})} \\
S0 (dust=N)          & 13  &  2.39$\pm$0.81  &  9.90  &  7.43$\pm$0.18   & 0.58 \\
S0 (dust=n,N)        & 17  &  2.70$\pm$0.77  &  9.96  &  7.57$\pm$0.18   & 0.61 \\
S                    & 25  &  2.27$\pm$0.48  & 10.09  &  7.18$\pm$0.15   & 0.52 \\
S0/Es,b (dust=Y)$^*$ & 15  &  2.70$\pm$0.77  &  9.96  &  6.11$\pm$0.18   & 0.52 \\
E/Es,e               & 24  &  2.00$\pm$0.25  & 11.32  &  8.84$\pm$0.14   & 0.36 \\
BCG$^*$              & 10  &  2.00$\pm$0.25  & 11.32  &  8.54$\pm$0.14   & 0.31 \\
\multicolumn{6}{c}{$\log(M_{\rm bh}/{\rm M}_\odot) = A[\log(M_{\rm
      *,gal}/\upsilon\,{\rm M}_\odot) - C] +B$ (Appendix~\ref{Apdx1})} \\
E/Es,e               & 24  &  2.06$\pm$0.26  & 11.35  &  8.85$\pm$0.15   & 0.37 \\
E/ES,e/S0(dust=Y)    & 39  &  2.27$\pm$0.25  & 11.27  &  8.69$\pm$0.11   & 0.43 \\
BCG$^*$              & 09  &  2.27$\pm$0.25  & 11.27  &  8.39$\pm$0.11   & 0.36 \\
\multicolumn{6}{c}{$\log(M_{\rm bh}/{\rm M}_\odot) = A[\log(M_{\rm
      *,gal}/\upsilon\,{\rm M}_\odot) - C] +B$ (Appendix~\ref{Apdx1})} \\
major mergers        & 48  &  2.07$\pm$0.19  & 11.37  &  8.83$\pm$0.10   & 0.41 \\
S                    & 25  &  2.68$\pm$0.46  & 10.79  &  7.17$\pm$0.14   & 0.52 \\
S (w/o Circinus)     & 24  &  3.04$\pm$0.59  & 10.81  &  7.21$\pm$0.14   & 0.47 \\
S0/Es,b (dust=Y)     & 15  &  2.65$\pm$0.54  & 11.08  &  8.30$\pm$0.20   & 0.52 \\
\hline
\end{tabular}

The sample size, $N$, of each galaxy type is given in Column~2.  The type
`major mergers' includes the BCG, the ordinary E/ES,e galaxies, and the S0
galaxies with dust = Y.  The relations are derived from a Bayesian analysis
that treats the data symmetrically rather than minimising the root mean square
(rms) scatter, $\Delta_{\rm rms}$ dex, about the fitted relation in the BH
direction. An asterisk ($*$) on the galaxy type indicates an estimated rather
than measured relation (see Figure~\ref{Fig1} and the Appendix~\ref{Apdx1}).
The $\upsilon$ term equals 1.0 here and whenever using stellar mass estimates
consistent with those derived from eq.~4 in \citet{Graham:Sahu:22a}.
\end{table}

Connected with the $B/T$ ratios, 
several $M_{\rm bh}$-$M_{\rm *,gal}$ relations are shown in Appendix~\ref{Apdx1}
As no galaxy decompositions are required, these may 
be more amenable for studying ensembles of BH mergers and the associated ocean
of gravitational waves they produce \citep{2015Sci...349.1522S,
  2021ApJ...917L..19G, 2022arXiv220306016A}.

\section{Discussion}

\subsection{Making tracks}

Figure~\ref{Fig1} reveals several $M_{\rm bh}$-$M_{\rm *,sph}$ scaling relations,
illustrating the march of galaxies to larger masses and
different morphological types.  
There are fitted relations for non-dusty S0
galaxies \citep{Graham-S0}, S galaxies
\citep{2019ApJ...873...85D,Graham:Sahu:22a}, and ordinary E/ES,e
galaxies (Eq.~\ref{Eq1}).  In addition is the trend for dusty S0 galaxies \citep{Graham-S0},
which is offset from the relation defined by non-dusty S0 galaxies, and the
trend for BCGs, which we have just seen is offset from the relation defined by
the non-BCG E/ES,e galaxies.  Collectively, 
the adjacent relations track a sequence of increasing chaos, albeit with spirals blooming
along the way.  The increasing entropy, 
revealed through the growth of `dynamically hot' spheroids at the expense
of ordered rotating discs, results in convergence towards pure E galaxies.
This evolution between and along the relations 
could be quantified with a chaos parameter, such as the $B/T$ stellar mass ratio 
or dynamical mass ratio: 
$\sigma^2R_{\rm sph}/V^2h_{\rm disc}$, 
where $R_{\rm sph}$ is a suitable radius for the spheroid, 
$V$ is the disc rotation at some outer radius,
and $h_{\rm disc}$ is the disc scale-length.

The relatively dust-poor S0 galaxies on the left-hand side of
Figure~\ref{Fig1} might be quasi-primordial if frozen in time due to a
cluster environment which stripped away their gas, eroded their satellites,
and inhibited galaxy mergers due to high-speed passages within the cluster
swarm.  In contrast, the dusty S0 
galaxies may be built from wet S$+$S mergers (lower blue arrow) and 
wet S0$+$S mergers (middle and upper blue arrows).  Such mergers can result in
gas clouds shocking against each other, falling inwards to perhaps 
form a new disc, and sparking 
galaxy-centric bursts of dusty star formation. 
The relatively cool atmospheres of asymptotic giant branch stars rich in
carbon and oxygen, or exploding supernovae, previously sprayed metals into the
interstellar medium.  Initially, most of these elements stay in a gas phase,
although some quickly condense into dust particles as the stellar winds/ejecta
expand and cool. These refractory dust grain cores can grow 
substantial mantles as they enter dense metal-enriched gas clouds within the cooling
interstellar medium \citep{2003ARA&A..41..241D}. 
Indeed, the high metallicity, high dust content, and high density of neutral
gas will aid the gas cooling, molecule formation, and cloud condensation.
A sequence of increasing dust-to-H{\footnotesize I} with metallicity, [O/H],
can be seen in, for example, \citet[][their figure~6]{2008ApJ...678..804E}. 
The result is a somewhat
distinct population of dusty, high-mass S0 galaxies.

The few dust-poor S0 galaxies overlapping with the S galaxies might be former
S galaxies \citep{2022MNRAS.513..389R} which have lost their dust, gas,
and spiral density wave due to entering a harsh cluster environment 
\citep{2010AJ....140.1814Y}.  
Their satellites may then effectively evaporate, 
thereby contributing to the 
intracluster light rather than building up (the bulge of) the central galaxy
\citep{2007ApJ...668..826C}.  
Such evolution, or rather stagnation, is often
expressed in terms of S galaxies fading to become S0 galaxies.

Finally, the grey and black arrows in Figure~\ref{Fig1} denote major
dry mergers, in which $M_{\rm *,sph}$ can increase more than $M_{\rm bh}$ if
some of the progenitor galaxies' disc stars get folded into the newly wedded
galaxy's spheroidal component. These arrows show transitions from S0 to E to BCG (the upper
set of arrows) and from S0 to ES,e to E to BCG (the lower set of arrows).  For
example, the black arrow pair above NGC~1322 reflects a major dry merger of
two S0 galaxies with a $B/T$ ratio of 0.5; the end
product represents a doubling of the BH mass and a quadrupling of the
spheroid stellar mass.
The merger remnant is an E galaxy if the orbital angular momentum cancels.
Should the net angular momentum of the system not cancel, then the system will
not make it across to the E sequence but fall short, forming either an ES
galaxy with an intermediate-scale disc or an S0 galaxy with a dominant
disc. Once a galaxy's stellar mass is great enough, at $M_{*} >$ $\sim
10^{11}$ M$_\odot$, it is not uncommon for galaxies to be immersed in a
million-degree corona, which destroys and removes the dust and cooler
star-forming gas \citep{2003ApJ...599...38B, 2003ARA&A..41..241D,
  2004tcu..conf..213D, 2012NJPh...14e5023M}.

\begin{figure}
\begin{center}
\includegraphics[trim=0.0cm 0cm 0.0cm 0cm, width=1.0\columnwidth, angle=0]{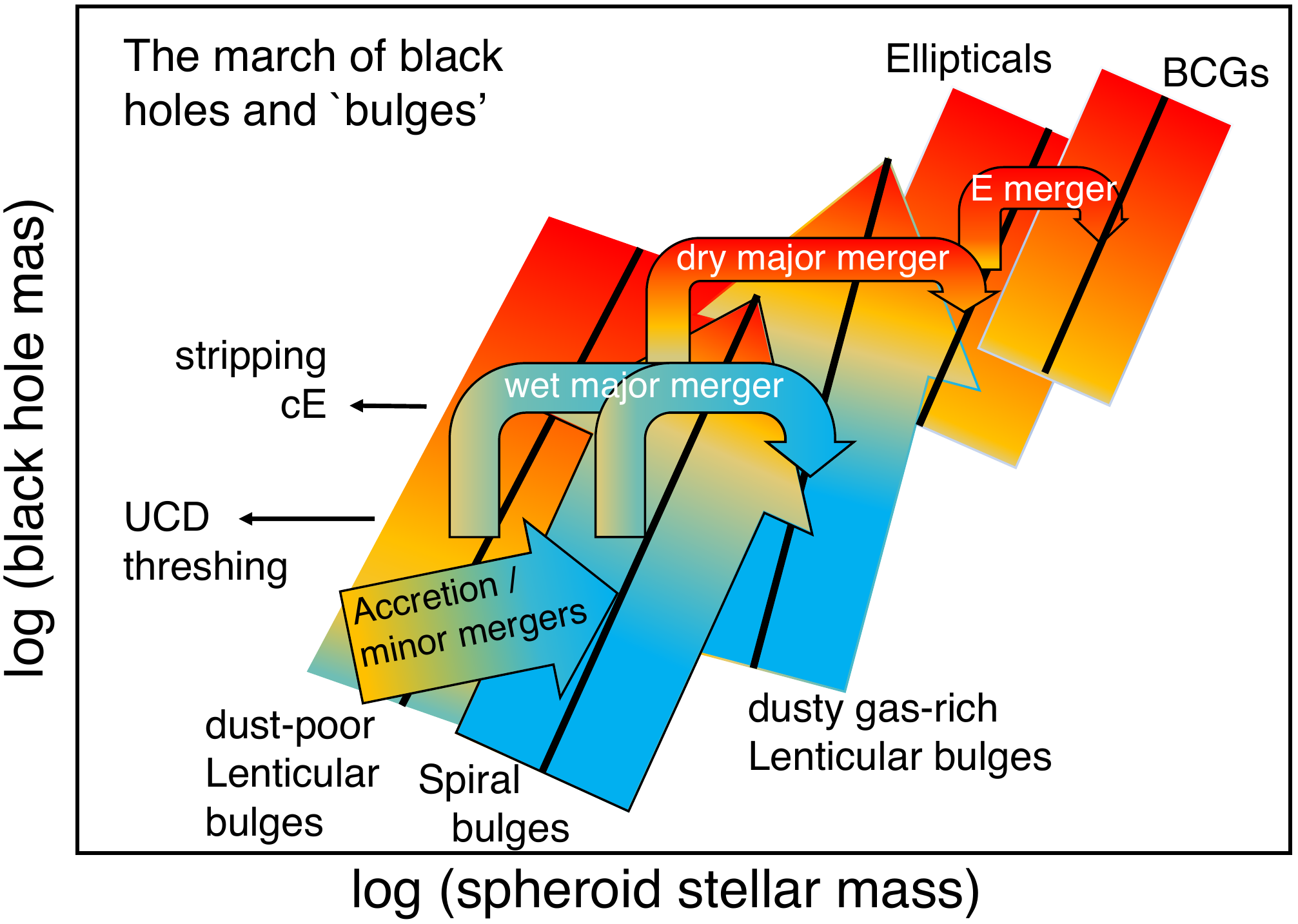}
\caption{Morphologically-aware $M_{\rm bh}$-$M_{\rm *,sph}$ schematic.  The
 progression of BH mass and `bulge' mass, i.e., the stellar mass of the
 spheroidal component of galaxies. 
 }
\label{Fig2}
\end{center}
\end{figure}

\begin{figure*}
\begin{center}
$
\begin{array}{cccc}
\includegraphics[trim=0.0cm 0cm 0.0cm 0cm, height=0.33\textwidth,
  angle=0]{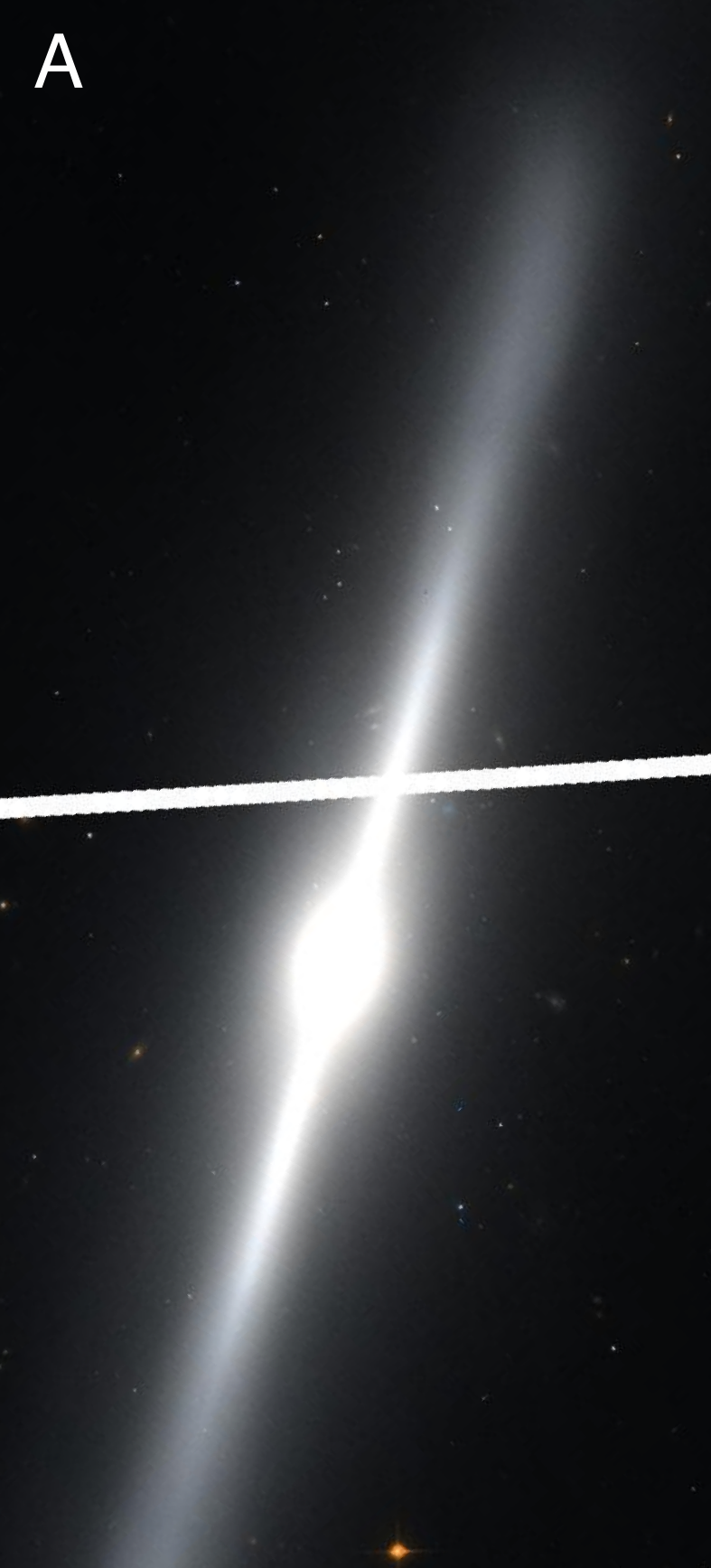} &
\includegraphics[trim=0.0cm 0cm 0.0cm 0cm, height=0.33\textwidth,
  angle=0]{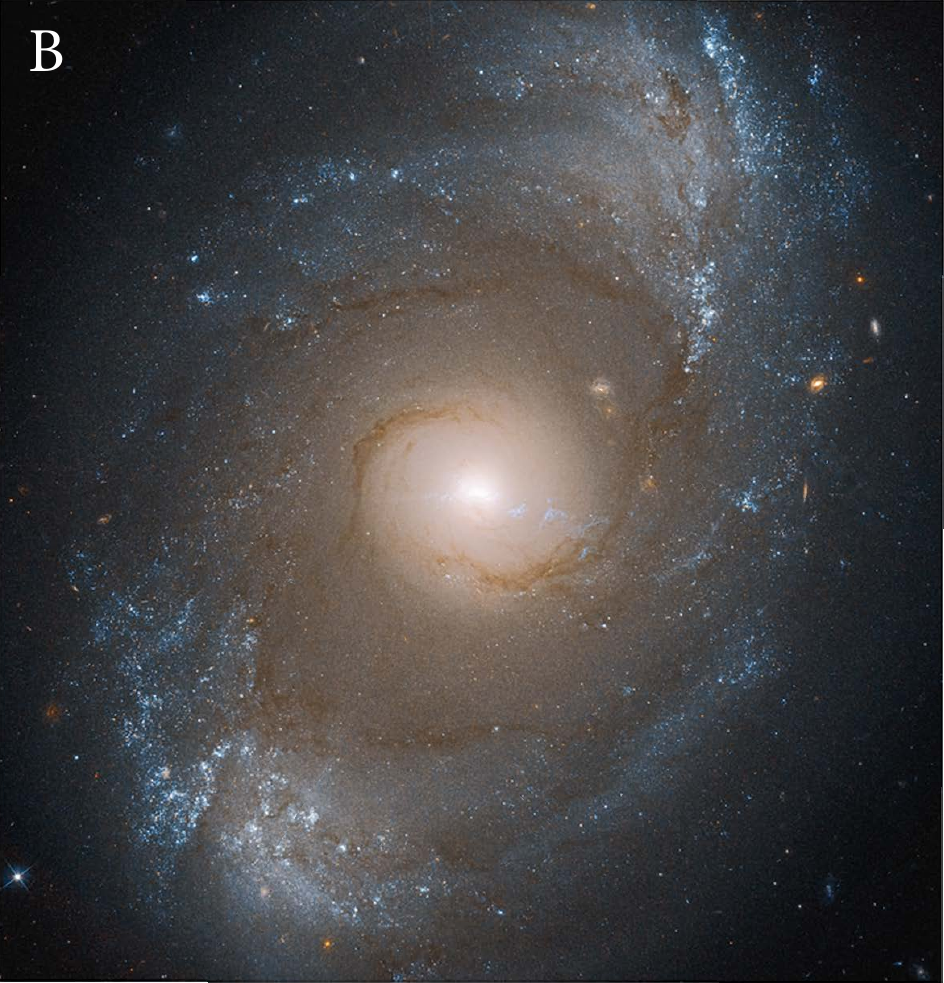} &
\includegraphics[trim=0.0cm 0cm 0.0cm 0cm, height=0.33\textwidth,
  angle=0]{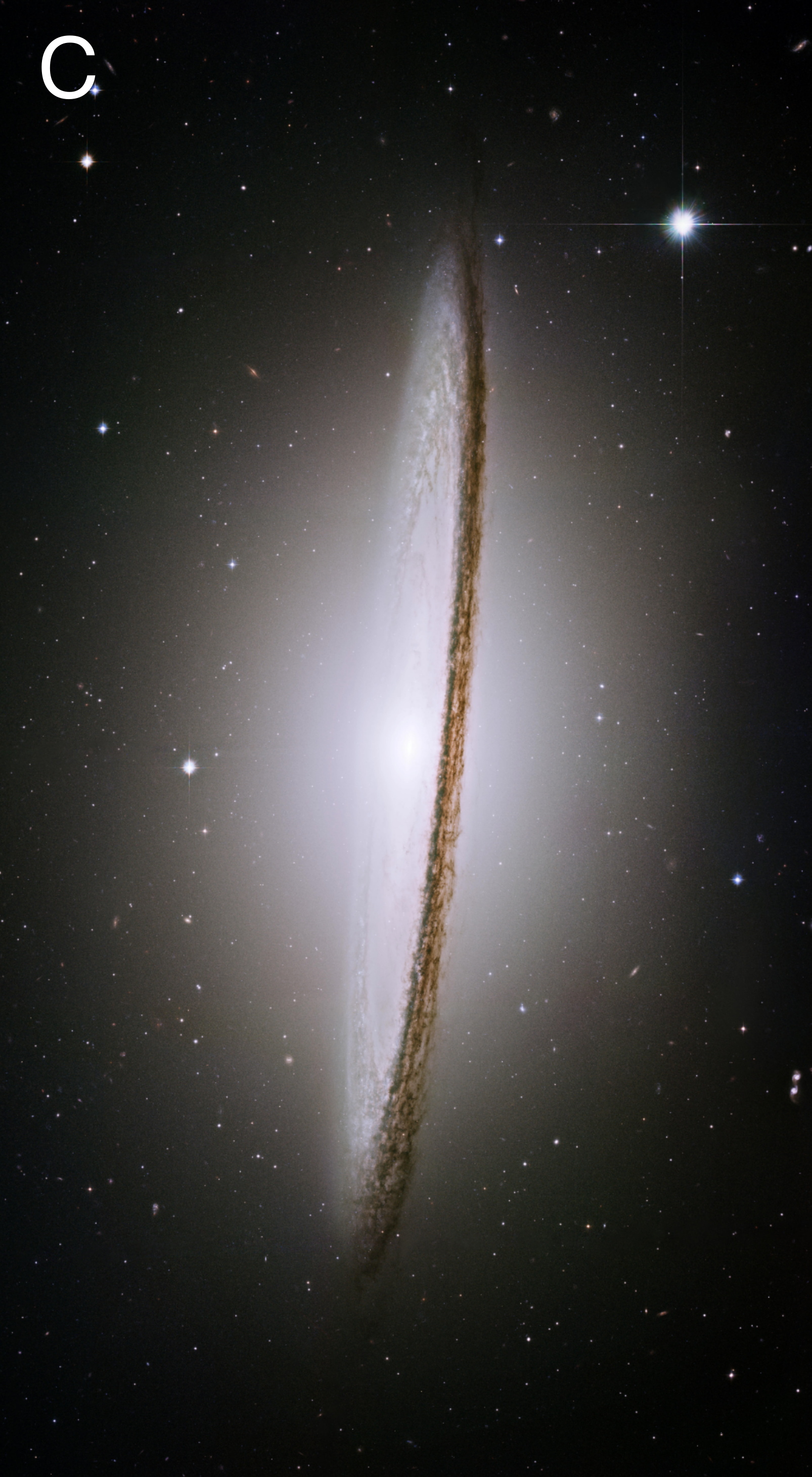} & 
\includegraphics[trim=0.0cm 0cm 0.0cm 0cm, height=0.33\textwidth,
  angle=0]{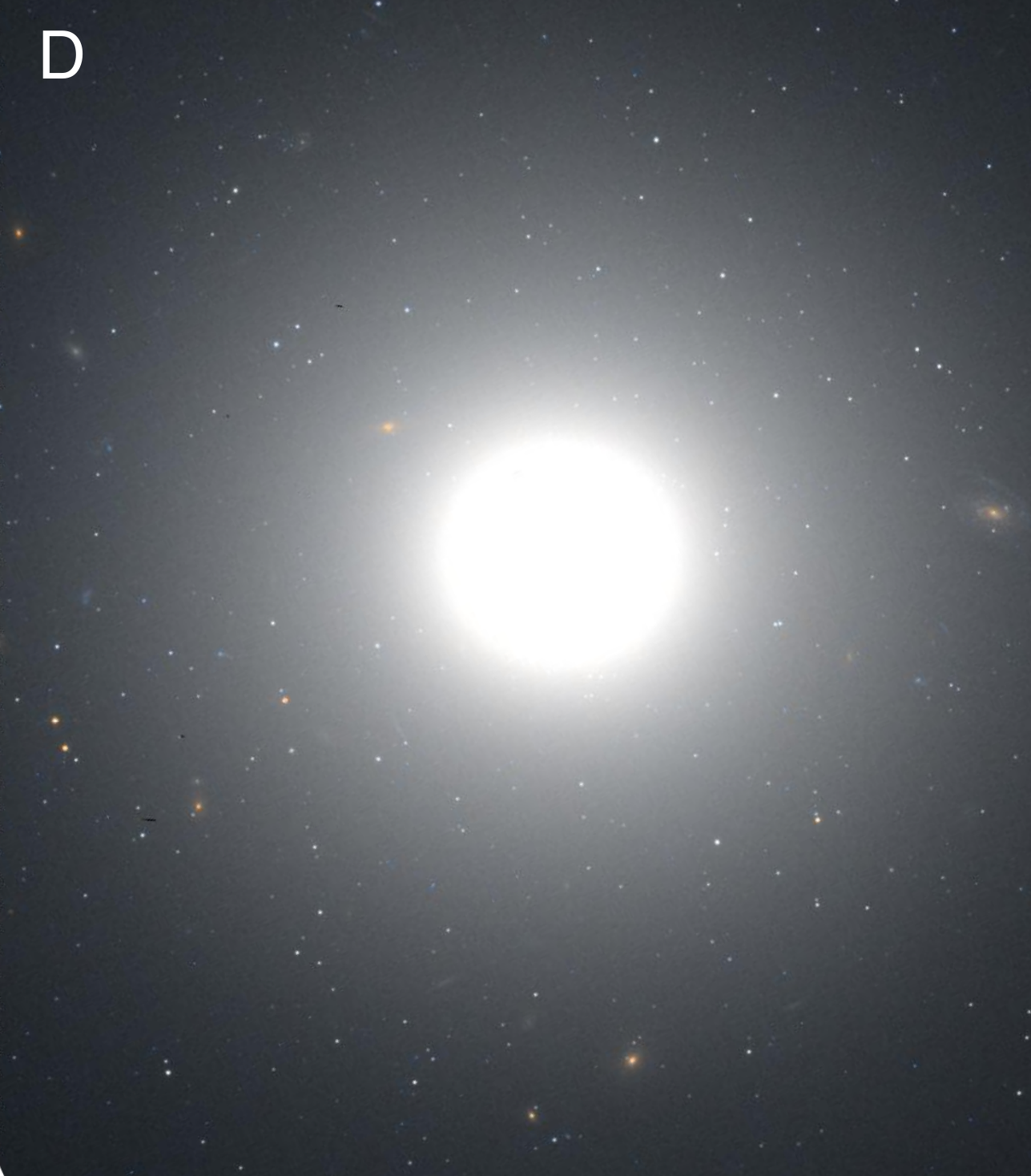} \\
\end{array}
$
\end{center}
\caption{Intergalactic speciation. 
Panel a) Dust-poor S0 galaxy NGC~4762 (HST Prop.\ 9401. PI.\ P.Cote. F850LP/F475W ACS/WFC).
Panel b) S galaxy NGC~4151 
(HST Prop.\ 13765. PI.\ B.Peterson. F814W/F350LP WFC3/UVIS.
STScI/NASA, ESA, Joseph DePasquale).
Panel c) Dust-rich S0 galaxy NGC~4594 (NASA and the Hubble Heritage Team. STScI/AURA).
Panel d)  E galaxy NGC~1407 (HST Prop.\ 9427. PI.\ W.Harris. F814W/F435W ACS/WFC). 
The white stripe in panel A is due to the camera join.
} 
\label{Fig3}
\end{figure*}


Overlaying the galaxies' morphological type onto the $M_{\rm bh}$-$M_{\rm
 *,sph}$ diagram has painted 
a new picture of the accretion history of galaxies.
This augmented scaling diagram reveals the major mergers, such as the
(i) transition from dust-poor S0 galaxies to dusty S0 galaxies built from
   wet mergers and having high $M_{\rm *,gal}/M_{\rm bh}$ ratios, 
(ii) the dry merger of S0 galaxies to produce ES and E galaxies, and 
(iii) the merger of E galaxies (and massive S0 galaxies) to produce BCG.
Furthermore, the $M_{\rm bh}$-$M_{\rm *,sph}$ diagram suggests 
that gas accretion and minor mergers onto dust-poor S0 galaxies have 
created the spiral galaxies.  
A cleaner representation of the data and tracks in Figure~\ref{Fig1} is
summarised through a simplified schematic in Figure~\ref{Fig2}. It captures
the major trends and transitions.  This is also shown pictorially in
Figure~\ref{Fig3}.  The S galaxy NGC~4151 shown there is a Seyfert with faint,
wispy arms extending beyond the displayed frame. It is reminiscent of UGC~6614
(not in sample), which has a low surface brightness disc with extended wispy
arms and a prominent bulge with an active galactic nucleus
\citep{1998AJ....116.1650S, 2006ApJ...651..853D}.  UGC~6614 is H{\footnotesize
  I}-rich and may have partially grown by accreting a dwarf galaxy
\citep{1997AJ....114.1858P}.

For the S galaxies, the spiral-arm winding angle correlates with the BH mass 
\citep{2008ApJ...678L..93S,2017MNRAS.471.2187D}, 
offering a view
of the changing S galaxy morphology along the S galaxy $M_{\rm bh}$-$M_{\rm
  *,sph}$ relation.  The low-mass S galaxies with low $B/T$ ratios (see 
Appendix~\ref{Apdx1}) tend to have
loosely wound spiral arms, while the opposite is observed at high masses.
The spiral patterns in S galaxies form in discs; that is, a disc is first
required.  The formation of the S galaxies, sandwiched between the dust-poor
and dust-rich S0 galaxies in the $M_{\rm bh}$--$M_{\rm *,sph}$ diagram 
suggests that gas accretion and minor 
mergers may be required for their emergence. 
A consequence is that our Milky Way galaxy was likely an 
S0 galaxy in the past before it merged with the {\it Gaia}-Sausage-Enceladus
satellite galaxy 10 Gyr ago.  However, if the mass ratio of merging galaxies is too close
to 1, the outcome may be more destructive, producing an S0 galaxy like
Centaurus~A. Indeed, such an outcome is expected when the Milky Way eventually
collides with the Andromeda galaxy in several Gyrs \citep{2012ApJ...753....9V, 2013JCAP...04..010E,
  2020A&A...642A..30S}.\footnote{Given the expected dusty nature of the merger product, a more
  apt name than ``Milkomeda'' may be ``Dustomeda''.}

Suppose the more massive of the original, currently dust-poor, S0 galaxies had
more satellites than the lower mass dust-poor S0 galaxies --- akin to the
increased numbers of dark matter subhalos observed in simulations
\citep{1999ApJ...524L..19M, 2013ApJ...767..146I}.  
In that case, then satellite capture and integration into the central S0
galaxy may build more massive bulges in the more massive dust-poor S0
galaxies.  Such growth may also contribute to the trend of the $B/T$ ratio
along the (Sa-Sb-Sc) S galaxy sequence \citep{2008MNRAS.388.1708G}.
and, in turn, contribute to a 
tightening of the winding of the spiral arms generated by the density waves
\citep{1964ApJ...140..646L}.  Such {\it harvesting} of satellites may help
complete the picture of galaxy evolution, explaining why Sa galaxies have
bigger $B/T$ ratios than the less massive Sc galaxies and perhaps partly
explaining the missing satellite problem \citep{2010ApJ...709.1138D}.
The merger-driven galaxy evolution revealed by the BH mass scaling diagrams
(Figure~\ref{Fig1}) should also aid our understanding of the 
(dark matter halo mass)-(galaxy stellar mass) 
relations as a function of galaxy type
\citep{2021A&A...650A.113B,2021A&A...649A.119P}.

Due to dynamical friction 
\citep{1943ApJ....97....1C, 1974SvA....18..180B,                             
  1975ApJ...196..407T, 2011MNRAS.416.1181I, 2014ApJ...785...51A},  
albeit with competing evaporative effects \citep{1989MNRAS.241..849O, 2017MNRAS.470.1729M}, 
the currently dust-poor S0 galaxies that once resided, and may still reside, 
in richer globular clusters system are expected to have 
imprisoned more globular clusters at their centre \citep{1993ApJ...415..616C}.  
They should, therefore, 
contain a more massive nuclear star cluster \citep{2014MNRAS.444.3738A, 2022MNRAS.516.4691L}. 
Unlike in gas-poor globular
clusters, BHs can feed and grow at the centres of galaxies
--- and perhaps rapidly so \citep{2011ApJ...740L..42D} ---, and a (black
hole)-(nuclear star cluster) mass relation exists \citep{2020MNRAS.492.3263G}.
This relation may have its origins in the dust-poor S0 galaxy sequence.

\subsection{Acquisitions and mergers}

Almost as soon as the first S galaxy was discovered
\citep{1850RSPT..140..499R}, it was suggested that a tidal encounter with
another galaxy might produce spiral-like `tidal arms'
\citep{Roche:1850, 1852AJ......2...95A, 1906PASP...18..111A,
  1951pca..conf..195H}.  
In addition, infalling perturbers may induce a
\citep[transient,][]{2018ApJ...853L..23B,2011MNRAS.410.1637S} spiral pattern
\citep{1966ApJ...146..810J,2008ASPC..396..321D, 2009ApJ...700.1896K} or a bar
\citep{2002NewA....7..155S} --- which may aid some longer-lasting grand-design
spirals --- and provide gas for ongoing star formation.
To date, the Milky Way galaxy appears to have had at least one significant
merger 10 Gyr ago, with the Gaia Sausage-Enceladus satellite
\citep{1999Natur.402...53H, 2018MNRAS.478..611B, 2018Natur.563...85H,
  2019NatAs...3..932G}, and perhaps an even greater merger before that
\citep{2021MNRAS.500.1385H}.  This is in addition to many increasingly lesser
  mergers involving, for example, the Sagittarius and Canis Major satellite
  galaxies \citep{2001ApJ...549L.199M,2019MNRAS.486.3180K}, and 
perhaps explaining Gould's Belt \citep{2009MNRAS.398L..36B}. 
Indeed, data from the ESA Gaia satellite has revealed that 
the disc of our Galaxy is notably unsettled \citep{2018Natur.561..360A,
  2018A&A...616A..11G, 2021MNRAS.504.3168B}.
Furthermore, disrupted dwarf or satellite galaxies are now routinely seen around S galaxies
\citep{2008ApJ...689..184M,2010AJ....140..962M,2016A&A...588A..89J,2021ApJ...907...85M}.

The extent to which the low-mass S0 galaxies have `harvested' systems from
their neighbourhood, and undergone star formation, could be substantial given
the factor of four difference in galaxy stellar mass seen between the
dust-poor S0 galaxies and the spiral galaxies at fixed BH mass 
(see Appendix~\ref{Apdx1}). 
However, there may have been considerable retardation of growth in the
dust-poor S0 galaxies located in clusters, and groups, due to a curtailed supply of gas and
stripping of stars \citep{1972ApJ...176....1G, 2008ApJ...672L.103K, 2009MNRAS.399.2221B}.  Furthermore, the
early-type spiral (Sa/Sb) galaxies with big bulges may have been built
by a major merger followed by disc-building 
\citep{2002NewA....7..155S,2009A&A...507.1313H}.  In contrast, the ES,b galaxies may not have
experienced the subsequent disc re-growth that these early-type S galaxies
did. 

It stands to reason that our Milky Way galaxy, perhaps first recognised as a
spiral system 170 
years ago \citep{1852AJ......2...95A}, see also \citet{1899ApJ.....9..149S}, was 
not born an S galaxy but was previously an S0 galaxy.  This conclusion
is supported by a myriad of stellar chemical and kinematic 
information \citep{2019ApJ...874L..35M, 2020A&A...636A.115D}.
The abundance of disc galaxies seen at high redshifts by 
\citet{2022ApJ...938L...2F} using the {\it James Webb Space Telescope} ({\it JWST})
also supports a picture of early disc galaxy formation.  
S galaxies would then consist of an old S0 galaxy disc, 
in which a thin disc has formed, and a spiral emerged
\citep{2017ApJ...850...61Y}, albeit with the ongoing competition with 
mergers (and spiral arms) which can dynamically heat and thicken a disc
\citep{1992ApJ...389....5T,2008ASPC..396..321D, 2018MNRAS.479L.108K}. 
It would be interesting to learn, through JWST observations, 
when the spiral patterns formed in the S0 galaxies, presumably
marking when a cold gas disc formed and a density wave emerged from the
differential rotation of the galaxy disc.

Cosmological simulations reveal an abundance of satellites contributing to the
growth of galaxies 
\citep[e.g.,][]{2017MNRAS.472.4343C,2021MNRAS.507.4211E,2022MNRAS.513.1867D}.  The star
cluster Nikhuli, in the disturbed S galaxy NGC~4424, is thought to represent the remains of
a disrupted dwarf early-type galaxy that may be delivering a BH into NGC~4424
\citep{2021ApJ...923..146G}, and potentially generating gravitational waves
should a second massive BH already reside there 
\citep{2007PhRvL..99t1102B,2009CQGra..26i4036M}.  Perhaps the Gaia
Sausage-Enceladus satellite stream also contained a migrant BH brought into our
Galaxy. 

\begin{figure*}
\begin{center}
\includegraphics[trim=0.0cm 0cm 0.0cm 0cm, width=0.66\textwidth,
  angle=0]{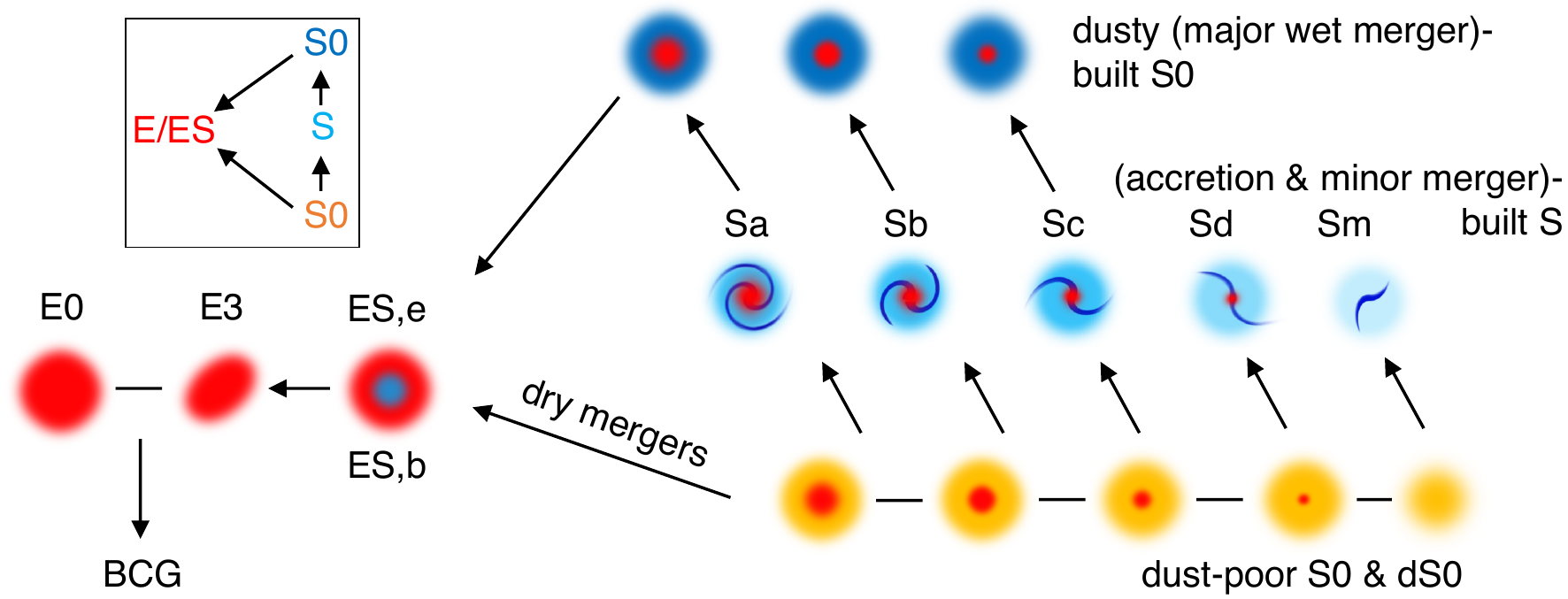} 
\caption{Not quite a regular three-pointed triangle,
  the `Triangal' (derived and simplified in Figure~\ref{FigS1}) reproduces elements of past galaxy
  morphology schemata \citep{1936rene.book.....H,1976ApJ...206..883V,2011MNRAS.416.1680C},
  recognises the ES galaxies, includes merger remnants, no longer
  presents the S0 galaxies as only a bridging population between the E and S 
  galaxies, connects the dwarf and dust-poor S0 galaxies with each 
  other, and reveals the accretion and merger-driven evolutionary paths
  between the morphological types. Spheroids are coloured red, while
  large-scale discs are
  coloured orange (old metal-poor/dust-poor), cyan (enriched with young metal-rich
  population), and blue (enriched/dusty).  The intermediate-scale discs in
  different ES galaxies have a range of stellar populations.} 
\label{Fig4}
\end{center}
\end{figure*}

In Figure~\ref{Fig4}, the dS0 galaxies are 
added to the low-mass end of the dust-poor S0 (not S) galaxy sequence, 
where one observes the spiralless disc galaxies with small bulges and 
$M_{\rm *,gal} < 10^{10} M_\odot$ (see Appendix~\ref{Apdx1}). 
If galaxies like IC~335, NGC~4452, and NGC~5866 (not in sample) 
are edge-on, near-bulgeless S0 galaxies, 
then some face-on examples may resemble 
low surface brightness galaxies, including ultra-diffuse galaxies 
\citep[UDGs:][]{1984AJ.....89..919S, 2017A&A...603A..18H}.  
They may follow the
curved size-(stellar mass) relation for early-type galaxies \citep{2019PASA...36...35G}
to lower masses and larger galaxy half-light radii.\footnote{In the absence of
  a centrally concentrated bulge, the galaxy size becomes the disc size.}
Some dS0 galaxies have been detected with faint spiral arms 
\citep{2000A&A...358..845J,2002A&A...391..823B,2003AJ....126.1787G}.  
Rather 
than being faded dwarf S galaxies, which are rare \citep{1984AJ.....89..919S}, 
they may be dwarf lenticular (dS0) galaxies 
attempting the transition to a late-type S galaxy 
\citep{2002ApJ...581.1039C,2017ApJ...840...68G} 
but retarded by lower numbers of satellites and reduced gas accretion.

\section{Placing developments in context}

One of the most well-known diagrams in astronomy is the `Hubble
sequence'
also known as the `Hubble tuning fork', reviewed in \citet{2019MNRAS.487.4995G}
along with other galaxy morphology schemata.  The sequence was initially
thought to be evolutionary, based on the `nebular hypothesis' from the
1700s. 
A legacy of that hypothesis is the `early-type' and `late-type' galaxy nomenclature
used over the past century for the E/S0 and S galaxies, respectively, plus the 
early- and late-type spiral galaxy designation \citep[e.g.,][p.867]{1925MNRAS..85..865L}, 
Originally, the Sa/Sb galaxies were thought to form before the Sc/Sd galaxies
\citep{1919pcsd.book.....J, 1921Obs....44..368R, 1925MNRAS..85.1014R}.  
The S0 galaxies were introduced later and positioned before the Sa galaxies, or rather,
between the E and Sa galaxies to give the sequence: (E0-E3)-S0-(Sa-Sc),
where the ellipticity of the E galaxies is denoted by $1-b/a$, with 
$b/a$ the observed axis ratio and thus apparently round 
galaxies labelled E0.   However, in terms of an 
increasing mass build-up, Figure~\ref{Fig1} (and Figure~\ref{FigS2}) reveal how one encounters the so-called
`late-type S' galaxies (Sc/Sd), known to have smaller BH masses, 
before the `early-type S' galaxies (Sa/Sb), known to have larger BH masses.
Although, within the `down-sizing' scenario \citep{1996AJ....112..839C}, 
the higher mass S galaxies might finish forming first. 

This section provides a brief overview of the significant advances which have
led to the emergence of a new, triangular-like galaxy sequence presented in
Figure~\ref{Fig4} and detailed further in Figure~\ref{FigS1}.  Dubbed the `Triangal', it 
reveals the morphological connections and, for the first time, the
merger-induced evolutionary pathways responsible for galactic speciation.
It identifies, and recognises the significance of, the dust-poor and dust-rich
S0 galaxies. 

Figure~\ref{Fig4} reflects that a galaxy's collisional record is evident from its
morphology.  As for the underlying, merger-driven, morphology-dependent 
$M_{\rm bh}$-$M_{\rm *,sph}$ scaling relations, BHs may now seem somewhat akin to
passengers, carried along by major mergers in which the redistribution of disc
stars, and the increase in orbital entropy, leads to the step-change
creation, i.e., `punctuated equilibrium', of more massive spheroids and the
transition to a new species of galaxy. 
Major wet mergers are, however, also associated with star formation and BH growth.  
Gaseous processes may be restricted 
to producing movement along the individual quadratic or steeper $M_{\rm bh}$-$M_{\rm *,sph}$ relations,  
yielding a kind of `gradualism' rather than evolution off any 
morphology-dependent relation.

Derived from the morphologically-aware $M_{\rm bh}$-$M_{\rm *,sph}$ diagram
(Figure~\ref{Fig1}), the schematic in Figure~\ref{Fig4} capture elements of
not just the
`Hubble sequence' but also the van den Bergh trident, introduced
within the Revised David Dunlap Observatory system \citep{1976ApJ...206..883V},
which was later re-expressed as the ATLAS$^{\rm 3D}$ comb
\citep{2011MNRAS.416.1680C}.  Since the van den Bergh trident was introduced,
there have been several significant developments, two of which trace back to
work by Sidney van den Bergh.  First was the realisation that many early-type
galaxies contain discs \citep{1990ESOC...35..231C,1990ApJ...362...52R} such
that low-luminosity\footnote{Absolute magnitude $M_B > -20$ mag, H$_0$=50 km
  s$^{-1}$ Mpc$^{-1}$.}  early-type galaxies are S0 galaxies rather than E
galaxies \citep{1990ApJ...348...57V}.  The abundance of rotating discs was
later witnessed through kinematic information
\citep{1998A&AS..133..325G,2011MNRAS.414..888E}.  Second was the realisation
that there are two subtypes of S0 galaxy: low- and high-luminosity
\citep{1990ApJ...348...57V}, with the origin of the high-luminosity S0 galaxies
now known to be due to wet mergers \citep{Graham-S0}.
These dusty high-luminosity S0 galaxies are not faded S galaxies 
\citep{1951ApJ...113..413S}
--- 
an idea which partly motivated the van den Bergh trident 
\citep{1976ApJ...206..883V} --- 
but S$+$S \citep{1998ApJ...502L.133B, 2003ApJ...597..893N, 2015A&A...579L...2Q} 
or S$+$S0 or (cold-gas rich but dust-poor) S0$+$S0 merger remnants. 
This accounts for why they are more massive 
than the S galaxy population \citep{1990ApJ...348...57V,2005ApJ...621..246B}. 
The sequence of dust-poor S0 galaxies seen in Figure~\ref{Fig1} are also not faded 
S galaxies but rather failed S galaxies that never were. 

The S0 galaxies are not the lynchpin they were initially thought to be
\citep{1925MNRAS..85.1014R,1936rene.book.....H}; 
that description seems more apt of the ES galaxies
\citep{1966ApJ...146...28L}, 
which are both `fast rotators' and `slow rotators', 
backtracking on themselves in the modified spin-ellipticity diagram for
galaxies \citep{2017MNRAS.470.1321B}.  
The ES galaxies are a bridging population
between the E and S0 galaxies, while the 
S galaxies are now a bridging population between the dust-poor
and dust-rich S0 galaxies.  

Recognising that S0 galaxies are not simply a single bridging population
(E0-E3)-S0-(Sa-Sc), 
nor are they a single low-(spiral strength) 
side to the disc galaxy distribution of $B/T$ ratio and spiral strength 
\citep{1976ApJ...206..883V}, alleviates a long-standing mystery.  While the
dusty S0 galaxies are a merger-built bridging population between the S and E
galaxies, the non-dusty S0 galaxies form both a low-mass extension of the E 
galaxies --- along a sequence of changing $B/T$ ratio and 
specific angular momentum \citep{1988A&A...193L...7B, 1992ASSL..178...99C}
 --- and provide a population 
for accretion, minor mergers, and the development of spiral 
structures.  

In addition to the $B/T$ ratio (Figure~\ref{FigS2}) --- known to correlate with the bulge 
mass \citep[e.g.][]{2008MNRAS.388.1708G} --- 
future work will explore the location 
in the $M_{\rm bh}$-$M_{\rm *,sph}$ diagram 
of S galaxies with different arm strengths \citep{1976ApJ...206..883V}. 
After checking if systems with weak/anaemic arms preferentially 
reside on one side of the S galaxy sequence, the location 
of disc galaxies with strong/weak/no bars 
\citep{1959HDP....53..275D} will be examined. 
Furthermore, it may be insightful to explore if other
features, such as ring-shape versus spiral-shape \citep{1959HDP....53..275D} 
or the wealth of fine detail captured by the
`Comprehensive de Vaucouleurs revised Hubble-Sandage' system 
\citep{2007dvag.book.....B}, occur in galaxies preferentially occupying a
specific part of the diagram.

\section*{Acknowledgments}

This paper is dedicated to the memory of Troy Charles Smith (1973 February
8--2023 February 2), a great friend and neighbour to interact with, and whose
``but what about...'' remarks would prompt one to query the status quo.
The author warmly thanks Drs Denis W.\ Coates and Vale Keith Thompson, 
formerly at Monash University, for past discussions.
Part of this research
was conducted within the Australian Research Council's Centre of Excellence
for Gravitational Wave Discovery (OzGrav) through project number CE170100004.
This work has used the 
NASA/IPAC Infrared Science Archive (IRSA), 
the NASA Extragalactic Database (NED), 
NASA's Astrophysics Data System (ADS) Bibliographic Services, 
and the Hubble Legacy Archive (HLA).

\section{Data Availability}

The data for this investigation consists of published \citep{Graham:Sahu:22a}
black hole masses, spheroid (and galaxy) stellar masses, along with the
galaxies' morphological type, including whether the E galaxies are BCG or
cD\citep{Graham:Sahu:22b}.  The spheroid masses are obtained from published
multicomponent decompositions, which separate bars and inner discs from the
spheroids\citep{2016ApJS..222...10S,2019ApJ...873...85D,2019ApJ...876..155S,Graham:Sahu:22b}.

\bibliographystyle{mnras}
\bibliography{Paper-BH-mass4}{}

\appendix

\section{The $M_{\rm bh}$-$M_{\rm *,gal}$ diagram}
\label{Apdx1}

\begin{figure*}
\begin{center}
\includegraphics[trim=0.0cm 0cm 0.0cm 0cm, width=1.0\textwidth, angle=0]{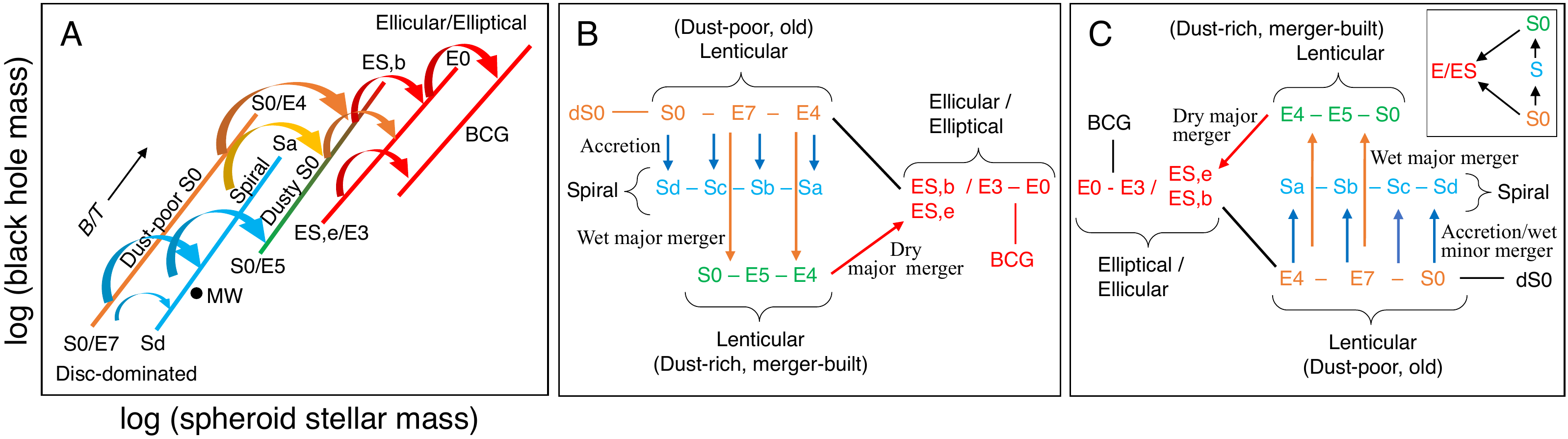}
\caption{Sequence of galaxy evolution. 
  Connections are revealed through the coevolution of galaxies
  and their BHs. The old E4 to E7 notation should be taken
  as representing S0 galaxies. The dusty S0 galaxies are coloured green due to
  their residence in the `green valley' within the galaxy colour-luminosity
  diagram \citep{2017ApJ...835...22P}. 
} 
\label{FigS1}
\end{center}
\end{figure*}

\begin{figure}
\begin{center}
\includegraphics[trim=0.0cm 0cm 0.0cm 0cm, width=1.0\columnwidth,
  angle=0]{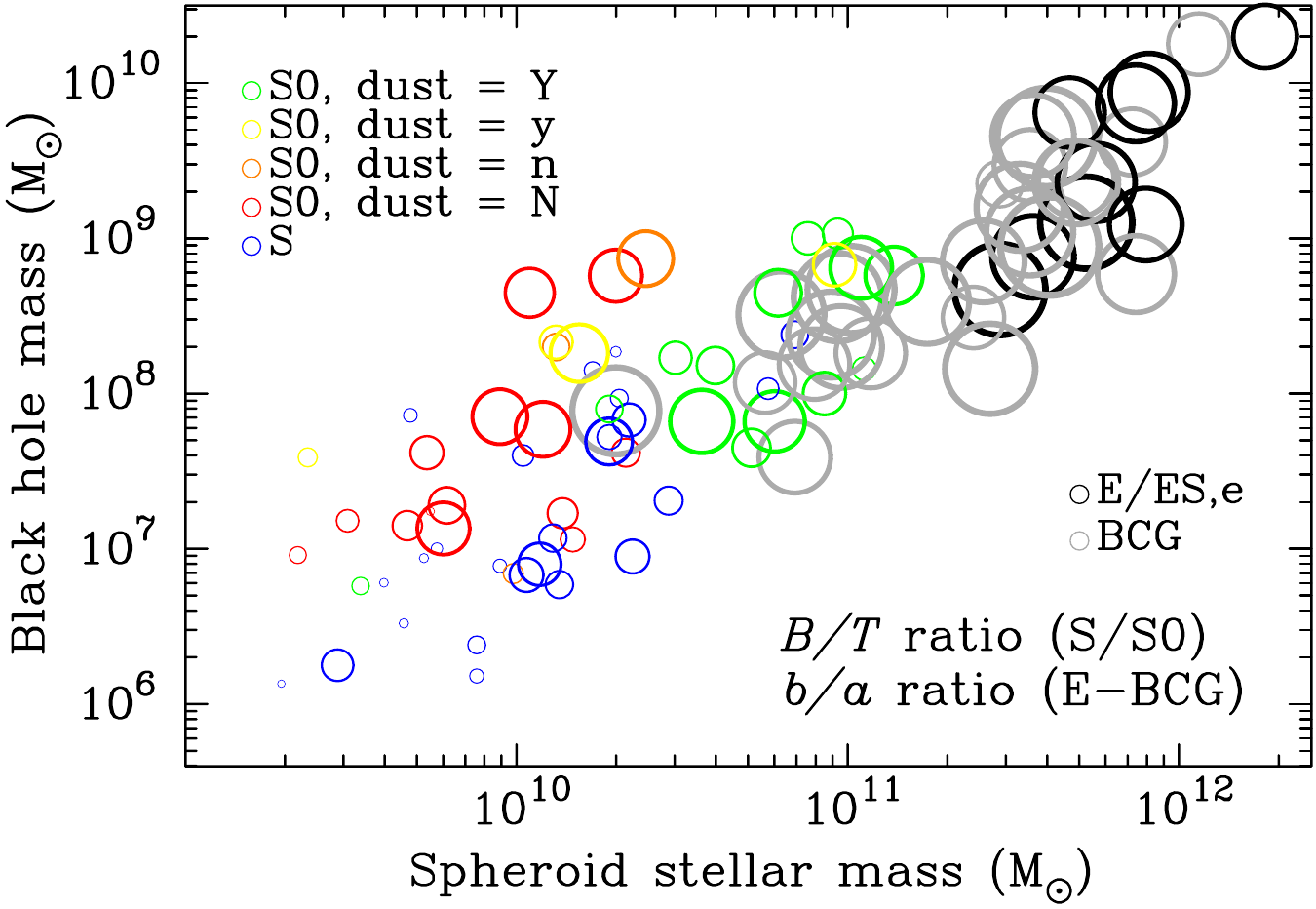}
\caption{Ratios in the $M_{\rm bh}$-$M_{\rm *,sph}$ diagram. 
The symbol size linearly reflects the $B/T$ 
ratios for the disc galaxies and the minor-to-major axis size ($b/a$) ratios for
the BCG, E, and ES,e galaxies.  The ratios are related to 
the specific angular momenta of a galaxy, the flattening and disc-dominance in
E/ES/S0 galaxies, and the S galaxy morphology.}
\label{FigS2} 
\end{center}
\end{figure}


\begin{figure*}
\begin{center}
\includegraphics[trim=0.0cm 0cm 0.0cm 0cm, width=0.8\textwidth, angle=0]{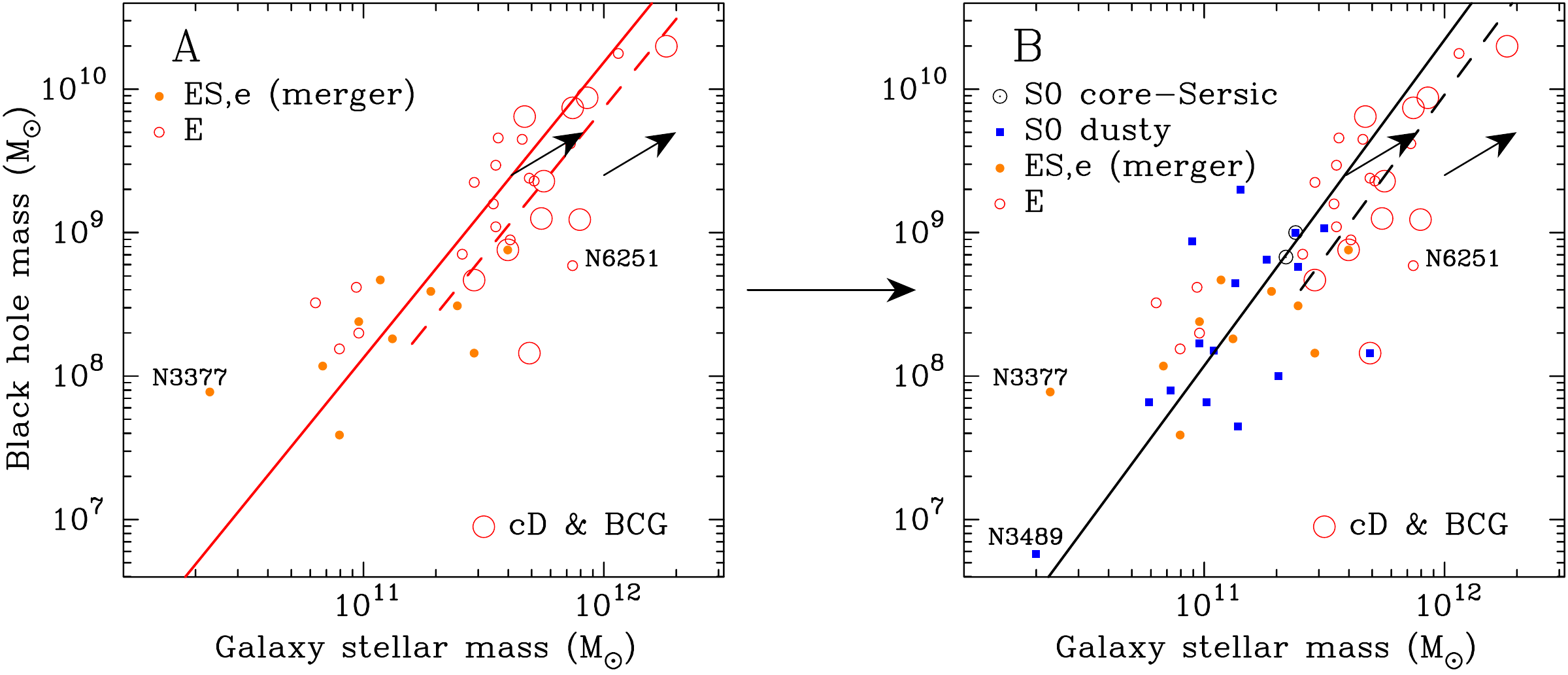}
\caption{Merger-built galaxies in the $M_{\rm bh}$-$M_{\rm *,gal}$ diagram. 
Left: Fit to 24 ordinary (non-BCG) E/ES,e galaxies (excludes the labelled galaxies). 
The BCGs \citep[listed in table~2 of][]{Graham:Sahu:22b} 
tend to reside to the right of the fitted relation, given in Table~\ref{Table1}.  Right:
Fit to 24 non-BCG E/ES,e galaxies plus 13 dusty S0 galaxies (including two
ES,b galaxies), and (two) core-S\'ersic S0 galaxies.  The plotted relation,
provided in Table~\ref{Table1}, has a logarithmic slope of 2.27$\pm$0.25.  The two dusty
galaxies most above the relation are the ES,b galaxies NGC~3115 and NGC~6861,
which have not grown a large-scale disc but were grouped with the S0 galaxies
due to their location at the top of the S0 galaxy $M_{\rm bh}$-$M_{\rm *,sph}$
relations.  The arrows have a slope of 1, representing the size of the jump
induced by a major dry merger capable of creating BCGs.  In both panels, the
dashed line represents the ensemble of such jumps and is thus offset by
$\log(2)=0.3$ dex in both axes from the solid line.  }
\label{FigS3}
\end{center}
\end{figure*}

Figure~\ref{FigS1} exposes the steps for extracting the `Triangal' (Figure~\ref{Fig4}) from the
$M_{\rm bh}$-$M_{\rm *,sph}$ diagram (Figure~\ref{Fig1}), revealing something like a
galactic family tree.
This section explores the $M_{\rm bh}$-$M_{\rm *,gal}$ diagram 
to further witness the galaxy connections, 
show that the separation of the dust-poor and dust-rich S0 galaxies is not a
consequence of the galaxy decomposition procedure, and 
establish which relations may exist without needing galaxy decompositions. 

The changes when switching from the $M_{\rm bh}$-$M_{\rm *,sph}$ diagram to
the $M_{\rm bh}$-$M_{\rm *,gal}$ diagram can be understood by looking at the 
$B/T$ ratios. 
For the disc galaxies, Figure~\ref{FigS2} reveals the varying $B/T$ ratios 
across the $M_{\rm bh}$-$M_{\rm *,sph}$ diagram and along the S0 and S 
sequences.  
For the ordinary E/ES,e galaxies and BCGs --- predominantly E
galaxies --- the 
semi-minor to semi-major axis size ($b/a$) ratios have been plotted 
instead of the $B/T$ ratios, which are close to 1. 
Ratios of 1.0 yield the largest circles in Figure~\ref{FigS2}. 

The offset nature of the BCGs relative to the ordinary (non-BCG) E/ES,e
galaxies seen in Figure~\ref{Fig1} can also be seen in the $M_{\rm bh}$-$M_{\rm 
  *,gal}$ diagram (Figure~\ref{FigS3}, panel A).  In passing, it is noted that
the galaxy sample's morphological types \citep{Graham:Sahu:22a,Graham:Sahu:22b}
confirm past allegations \citep{1966ApJ...146...28L,1970SvA....14..182G} that
most E4 and all E5, E6 and E7 designations were assigned to misclassified S0
galaxies.  Given the dominance of the spheroid in BCGs and E/ES,e galaxies,
the situation in panel A of Figure~\ref{FigS3} is similar to that seen in
Figure~\ref{Fig1}.  The ordinary E/ES,e galaxies follow a near-quadratic relation given
in Table~\ref{Table1}.

In panel B of Figure~\ref{FigS3}, 
the two merger-built core-S\'ersic S0 galaxies, 
the two dusty ES,b galaxies, 
and the dusty (non-BCG) S0 galaxies built from major wet mergers, have been
combined with the ordinary E/ES,e galaxies. 
Excluding the three labelled galaxies (NGC~3489, NGC 3377, and NGC~6251), 
these 39 non-BCGs give a relation with a logarithmic slope of 2.27$\pm$0.25 (see Table~\ref{Table1}), 
shown by the solid line in panel B of Figure~\ref{FigS3}.  
Additionally excluding the two dusty ES,b galaxies changes the logarithmic
slope to 2.30$\pm$0.25 and the intercept to 8.63$\pm$0.11.

In understanding the high-mass end of the $M_{\rm bh}$-$M_{\rm *,gal}$
diagram, the arrows in Figure~\ref{FigS3} show the pathway of dry, equal-mass mergers,
which shift systems off the steeper-than-linear $M_{\rm bh}$-$M_{\rm *,gal}$
relation defined by the ordinary (pre-merged) E/ES.e galaxies.  One can
appreciate how the BCGs have evolved away from this relationship with a
logarithmic slope of $\sim$2 by individually following a path with a slope of
1.  Due to the different pre-merger starting points in the $M_{\rm
  bh}$-$M_{\rm *,gal}$ diagram, the collective merger-induced shifts can
result in the BCG following an offset distribution which roughly preserves the
original logarithmic slope of $\sim$2. Greater numbers of BCGs will help to
quantify their distribution better.\footnote{The Bayesian analysis produces an
  $M_{\rm bh}$-$M_{\rm *,sph}$ relation for the BCGs with an uncertain 
  logarithmic slope of 2.89$\pm$1.12.  A higher logarithmic slope of 2.65-3.0 may
  signal the contribution from BCGs built from dusty S0/S galaxy mergers.}

\begin{figure*}
\begin{center}
\includegraphics[trim=0.0cm 0cm 0.0cm 0cm, width=0.66\textwidth, angle=0]{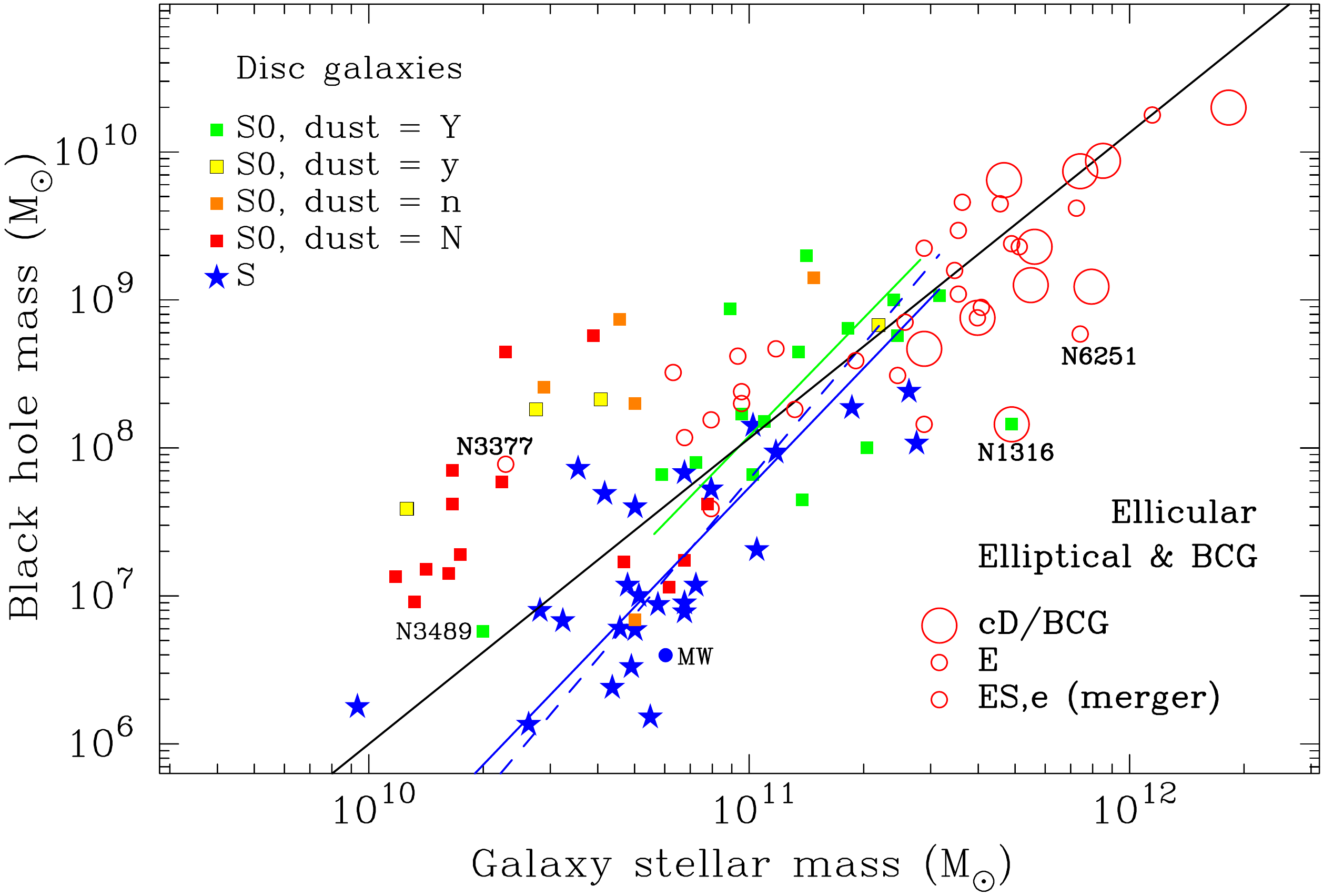}
\caption{$M_{\rm bh}$-$M_{\rm *,gal}$ diagram.  Similar to Figure~\ref{FigS3}
  but now showing all the galaxies (cf.\ Figure~\ref{Fig1}).  The black line, with a
  logarithmic slope of $2.07\pm0.19$ defined in Table~\ref{Table1}, stems from a fit to
  nine cD/BCG plus 24 non-BCG E/ES,e galaxies, 14 dusty S0 galaxies (including
  two ES,b galaxies and one core-S\'ersic S0), and a second merger-built
  core-S\'ersic S0 galaxy but with dust = y rather than dust = Y.  The labelled
  galaxies are excluded from the fit, as are the S galaxies and the S0
  galaxies without much dust, i.e.\ dust = N, n, or y.  This relation is defined
  by early-type galaxies having $M_{\rm *,gal} >$ (0.6-0.8)$\times 10^{11}$
  M$_\odot$ and is subject to sample (morphological sub-type) selection, as
  noted in the text.  The solid blue line comes from a fit to the S galaxies,
  and the dashed blue line was obtained upon excluding the lowest mass S
  galaxy, Circinus, see Table~\ref{Table1}. }
\label{FigS4} 
\end{center}
\end{figure*}

The sample size of nine BCG (excluding NGC~1316) is too low to obtain a
reliable fit, with the Bayesian analysis giving an 
uncertain logarithmic slope of 3.24$\pm$1.38.
Therefore, the dashed lines in Figure~\ref{FigS3} are not a fit but simply a shift by 
a factor of two to a higher BH and galaxy stellar masses.  
Adding these nine BCG to the other galaxies built from major mergers yields the relation 
\begin{equation}
\log(M_{\rm bh}/M_\odot)= 
(2.07\pm0.19)[\log(M_{\rm *,gal}/M_\odot)-11.37]+(8.83\pm0.10), 
\label{Eq2}
\end{equation}
shown by the solid line in Figure~\ref{FigS4}.\footnote{Excluding the two dusty ES,b
  galaxies changes the logarithmic slope to 2.13$\pm$0.18 and the intercept to
  8.79$\pm$0.10.}  While this is not too dissimilar from the relation defined
by the ordinary E/ES,e galaxies in panel A of Figure~\ref{FigS3}, it should 
be remembered that eq.~\ref{Eq2} depends on the relative number of morphological
types in the sample.  For example, a sample dominated by BCG would yield a
relation with a lower $M_{\rm bh}/M_{\rm *,gal}$ ratio at a given mass than a
sample dominated by non-BCG E/ES,e galaxies.  Although the extension of eq.~\ref{Eq2} to low
masses is shown in Figure~\ref{FigS4}, strictly speaking, eq.~\ref{Eq2} applies, modulo the above caveat, for 
early-type galaxies with $M_{\rm *,gal} >$ (0.6-0.8)$\times10^{11}$ M$_\odot$.

Given the scatter in Figure~\ref{FigS4}, robust regression lines were not
obtained for the dust-poor S0 galaxies. However, the trend of higher $B/T$
ratio with increasing mass (Figure~\ref{FigS2}) must generate steeper $M_{\rm
  bh}$-$M_{\rm *,gal}$ relations than $M_{\rm bh}$-$M_{\rm *,sph}$ relations
for each disc type. Thus the dust-rich (dust=Y) S0 galaxies should follow a
steeper relation than the near-quadratic relation shown in Figure~\ref{FigS4}.
Indeed, the S galaxies are best fit with a logarithmic slope of 2.68$\pm$0.46,
or 3.04$\pm$0.59, upon excluding the lowest mass S galaxy, Circinus.  This
latter slope is consistent with past results \citep{2018ApJ...869..113D}.  The
dust-rich S0 galaxies are best fit with a relation having a slope of
2.65$\pm$0.54 (see Table~\ref{Table1}), implying that the estimated slope of
2.70$\pm$0.77 in the $M_{\rm bh}$-$M_{\rm *,sph}$ diagram may be more reliable
at the lower end of the range.

Figure~\ref{FigS4} shows a 
tightening distribution when traversing from low to high masses.
For instance, below $\sim 10^{11}$ M$_\odot$, relatively dry mergers of galaxies above the
near quadratic relation (black line) 
will progress systems along a shallow path with a slope of $\sim$1
to reach the relation. 
In contrast, wet mergers of S galaxies below the relation must proceed along a
path with a logarithmic slope steeper than $\sim$2 in order to generate the dust-rich (dust=Y) S0 galaxies.
As such, in gas-rich systems, BH growth must out-pace stellar growth. 

Without a significant second round, or extended bout, of star formation
fuelled by metal-enriched gas \citep[e.g.,][and references
  therein]{2009ApJ...692L..24B}, it is expected that the low-mass, dust-poor
S0 galaxies will have a different stellar population and perhaps also a
different abundance of planets \citep{2012A&A...545A..32A} to the S and
dust-rich S0 galaxies.  Although, all should share the same old foundational
population.
Separating the dusty and non-dusty S0 galaxies may aid 
studies of globular cluster systems in and 
around galaxies.  For example, a prevalence of bi- or tri-modal populations of
globular clusters in the dusty (gas-rich merger)-built S0 galaxies 
could reaffirm the evolutionary picture
\citep{2001AJ....121.2974L, 2010MNRAS.404.1203F, 2013MNRAS.428..389P}, especially if the dust-poor
S0 galaxies are shown to display a uni-modal distribution of old, metal-poor,
blue globular clusters while the dusty S0 galaxies additionally have more
centrally concentrated metal-rich red globular clusters that formed during the
galaxy merger. 
A comparison with the (dry merger)-built
E galaxies may also be insightful for addressing how often the
intermediary stage of ES and dusty S0 galaxies is bypassed in the
evolutionary chain.

The relative dominance of thick and thin stellar discs from dust-poor S0
galaxies to S galaxies to dust-rich S0 galaxies might also hold important
details to the evolutionary chain revealed through the `Triangal'.  
Stellar discs might be born hot from turbulent gas clouds 
\citep[in situ formation:][]{2004ApJ...612..894B}. 
Thin (and thick) stellar discs can be `dynamically
heated' and thickened from minor mergers \citep{1993ApJ...403...74Q}, which
can additionally directly deposit accreted stars into a thick(er) disc
\citep{2003ApJ...597...21A, 2014MNRAS.443..828C}, while the gas from such mergers and direct
accretion can cool, settle to the mid-plane, and build a new thin disc of stars
within which a spiral density wave might later form
\citep{1964ApJ...140..646L}.  Gravitational perturbations from infalling
satellite galaxies may also trigger the formation of a transient spiral pattern
\citep{1966ApJ...146..810J,2013ApJ...766...34D}.
In passing, it is interesting to note that some of the more major wet mergers,
spawning dusty galaxies like 
Markarian 463 \citep{1989AJ.....97.1306H}, 
NGC 6240 \citep{2003ApJ...582L..15K}, 
and others \citep{2019ApJ...882...41H}, 
which are still in the early stage of merging, will contain two BHs for 
hundreds of millions of years prior to their coalescence.

The `Triangal' disfavours the `monolithic collapse' scenario
\citep{1962ApJ...136..748E} for E, S, and (now) dusty S0 galaxies. 
It suggests that the now-dust-poor S0 galaxies might be the only massive
systems which formed that way.  
That is, these would be primordial galaxies.
Due to `down-sizing', the most massive now-dust-poor S0
galaxies, 
with $M_{\rm *,sph}\sim$(1-2)$\times10^{10}$ M$_\odot$ and 
$M_{\rm *,gal}$ of a few $10^{10}$ M$_\odot$, may emerge early
\citep{2023ApJS..265....5H}. 
Early major mergers in the Universe 
would shift these galaxies from the top-end of the now dust-poor S0 galaxy sequence to
the top-end of the dust-rich S0 galaxy sequence, where some ES,b and
core-S\'ersic S0 galaxies reside.  
It is suggested here that the  
$10^{11}$ M$_\odot$ `red nuggets' at redshifts of 2 to 3 
\citep{2005ApJ...626..680D, 2011ApJ...739L..44D} might have formed from such
mergers. An early example may be reported in \citet{Labbe2023}. 

The `Triangal' reveals substantial orbital 
angular momentum in the Universe's first galaxies, and supports the unification
of dwarf and ordinary `early-type' galaxies \citep[][and references therein]{2019PASA...36...35G}.
The new galaxy sequence, encapsulated by the `Triangal' --- itself informed by
the low $M_{\rm bh}/M_{\rm *,sph}$ ratios in (dry merger)-built E galaxies,
core-S\'ersic galaxies and BCGs which experienced `galforming'\footnote{Some BCG
  and cD galaxies are expected to have such large depleted cores that a low
  \citet{1963BAAA....6...41S} index function will be more practical than the 
  core-S\'ersic function \citep{2003AJ....125.2951G}.} 
\cite{Graham:Sahu:22b}, plus the S galaxies rained down upon by satellites and
significant neighbours --- suggests a new hybrid galaxy evolution model.
While monolithic collapse may form the (now) dust-poor S0 galaxies, accretion
and major mergers eat away at the stellar orbital angular momentum, with
rising entropy building galactic bulges and spheroids.  With smaller galaxies
taking longer to come in from the cold, galactic speciation may be expected to
first occur in the more massive systems due to the reduced mutual
gravitational attraction between lower mass galaxies.

The `Triangal' can also help paint detail onto 
semi-analytic simulations \citep{2013MNRAS.433.2986W}, 
hierarchical models of galaxy evolution \citep{1978MNRAS.183..341W}, and 
cosmological $N$-body simulations that track mass growth but are devoid of
morphological information.  
The trends in Figure~\ref{Fig1} and Figure~\ref{FigS4} partly 
support the galactic metamorphosis 
seen in dark matter simulations that include gas and star 
formation  \citep{2002NewA....7..155S}.  However, this requires some qualification and
can be regarded as `hierarchical merging' with a couple of essential 
twists on the popular story.  Galaxy evolution appears to differ from 
the headline in which S galaxies merge to form an E galaxy, 
which either loses its gas to become `red and dead' 
\citep[][their figure~1]{2008ApJS..175..356H}
or experiences accretion to build a new disc 
\cite{2009Natur.457..451D} and possibly become an S0 or S galaxy 
While these things may happen, 
the modified picture presented here recognises that S galaxy mergers 
(i) form an S0 rather than an E galaxy \citep{2002MNRAS.333..481B, 2003ApJ...597..893N} and 
(i) retain their gas. Furthermore, the more common 
sequence appears to be S0 to S to dust-rich S0 to E galaxy. 
This offers some insight into why \citep{1976RC2...C......0D} assigned both
dusty Irregular galaxies \cite[Irr
 II:][]{1950MeLuS.128....5H} and S0/a galaxies into the same T=0 numerical
stage index.  The Irregular galaxies had previously been positioned at the end
of the galaxy sequence, after the late-type Sd/Sm spiral galaxies.  

Furthermore, the `Triangal' --- which embraces
elements of and captures evolutionary growth sequences not embodied in the
`Hubble sequence' or the van den Bergh trident
\citep{1976ApJ...206..883V,2011MNRAS.416.1680C} --- reveals how dwarf-mass
early-type galaxies are related to both ordinary (dust-poor) early-type
galaxies and low-mass late-type S galaxies.\footnote{The bulgeless and
  near-bulgeless dwarf early-type (dS0) galaxies are not displayed in the
  classic $B/T$ versus morphology diagrams
  \citep{1970ApJ...160..811F,2016ApJ...831..132G} due to sample selection
  avoiding or missing these low-mass galaxies.}  An emerging version of this
connection can be seen in the extension of the van den Bergh trident to low
masses \citep{2012ApJS..198....2K}.  However, that extension did not recognise
that there are two S0 galaxy sequences, that the dS0 galaxies only
directly connect with one of these two sequences, or that the spirals are a
bridging population between the S0 galaxies.
Moreover, the alleged disjointedness
\citep{2009ApJS..182..216K,2012ApJS..198....2K} between the dwarf and ordinary
early-type galaxies has been addressed \citep{2013pss6.book...91G, 2019PASA...36...35G}.

\section{Outliers}
\label{Apdx2}

Possible measurement errors aside, it is not surprising to find a small number
of galaxies deviating from the general trends.  Infrequent 
events, which may be worthy of further investigation, could spawn such
behaviour.  In what follows, a handful of outliers have been identified for the
sake of such future work.  The bulk of these galaxies have previously been
noted as outliers in specific scaling diagrams, and sometimes a plausible reason 
has already been identified.  

While the current figures do not display error
bars due to the apparent clutter/confusion when displaying many galaxy types,
they can be seen in \citet{Graham:Sahu:22a}.

\subsection{(Non-dusty) S0 galaxies}

In constructing the dashed orange line in Figure~\ref{Fig1}, the 
S0 galaxies with no visible dust or only a small nuclear dust ring or
dust disc were regarded as non-dusty S0 galaxies. 
However, the non-dusty S0 galaxy NGC~4342 was excluded because 
it has been stripped of its stars by an unknown amount
\citep{2014MNRAS.439.2420B}. 
Although the non-dusty S0 galaxy NGC~7457 was retained, it 
was previously flagged for having an unusually low `velocity dispersion' for its
BH mass \citep{2019ApJ...887...10S}.  With cylindrical rotation about its major
axis, it is a likely merger remnant \citep{2002ApJ...577..668S}. 
The only other non-dusty `S0 galaxy' to be excluded was the outlying galaxy NGC~1332, with 
its thin nuclear dust disc. NGC~1332 is an ES,b galaxy, which are treated as
S0 galaxies in the $M_{\rm bh}$-$M_{\rm *,sph}$ diagram.

\subsection{S galaxies}

Residing in the large Eridanus Group, 
the H{\footnotesize I}-deficient S galaxy NGC~1300 
 \citep{2005JApA...26...71O,2022ApJ...927...66W}
is the only S galaxy with a BH mass measurement and a
(3.6 $\mu$m)-derived spheroid stellar mass that was excluded.  
It has either an unusually low spheroid mass or a high BH mass 
relative to the S galaxy $M_{\rm bh}$-$M_{\rm *,sph}$ relation
\citep{Graham-S0}. 

Two bulgeless S galaxies (NGC 4395 and NGC 6926) had to be excluded due to
their absence of a spheroid, and NGC~5055 still needs to have its BH mass
measured.  Two formerly S galaxies were reclassified as S0 galaxies: NGC~2974
and NGC~4594 (the Sombrero galaxy shown in Figure~\ref{Fig3}).\footnote{As noted
  by \citet{2012MNRAS.423..877G}, NGC~4594 is sometimes described as an S
  galaxy within an E galaxy, as is NGC~5128
  \citep[Centaurus~A:][]{1999A&A...341..667M}, perhaps conforming with the
  major wet merger scenario.  If spiral arms are present in NGC~4594 --- although
  \citet{2015ApJS..217...32B} dispute any are --- they closely resemble two rings
  \citep{1984A&A...141..309B} with a reported winding angle of just 5 degrees
  \citep{2017MNRAS.471.2187D}.} 
This adjustment left a working sample of 25 S galaxies.

The S galaxies Circinus and NGC~2960, known to have experienced
separate merger events, were retained, as was NGC~4945, 
the S galaxy with the lowest $M_{\rm bh}$ and $M_{\rm *,sph}$. 
NGC~4945 was previously flagged as an outlier in
the $M_{\rm bh}$-(S\'ersic $n$) and $M_{\rm *,sph}$-$\sigma$ diagrams
\citep{2020ApJ...903...97S,Graham-sigma}.

\subsection{Dusty S0 galaxies (wet mergers)}

The $M_{\rm bh}$-$M_{\rm *,sph}$ relation for the dusty S0 galaxies is
currently represented by a shift of $\Delta M_{\rm *,sph}=\log(3.5)$ from the
relation for non-dusty S0 galaxies \citep{Graham-S0}.  As such, no dusty S0
galaxies are excluded from the $M_{\rm bh}$-$M_{\rm *,sph}$ diagram.  However,
the dusty S0 galaxies were used in the current paper to construct the $M_{\rm
  bh}$-$M_{\rm *,gal}$ relation for merger-built non-BCG galaxies
(Figure~\ref{FigS3}).  The dusty S0 galaxies NGC~404 (outside of the
plotting range in the figures) and NGC~3489 were excluded due to their
location at the extremity of the distribution and, thus, their potential to bias
the result defined by the other galaxies.  NGC~404 ($\log(M_{\rm
  bh}/M_\odot)=5.74\pm0.1$ dex, $\log(M_{\rm *,gal}/M_\odot)=9.19\pm0.16$ dex)
can be seen in figure~5 of \citet{Graham:Sahu:22a}.  It may be experiencing a
slight rejuvenation of its stellar population  \citep{2010ApJ...714L.171T}, or 
perhaps it is still experiencing
the tail of star formation from a more substantial merger event
\citep{2013ApJ...772L..23B}.

In Figure~\ref{FigS4}, 
the dusty S0 galaxies were used in the non-BCG merger-built sample, which involved 
adding the E and ES,e galaxies 
plus the two core-S\'ersic S0 galaxies in the sample. 
Excluded from this  more extensive set were the ES,e galaxy NGC~3377, the E galaxy
NGC~6251, and the S0 galaxy NGC~3489.

\subsection{Ordinary (non-BCG) ES and E galaxies}

The ES,e galaxy NGC~3377 has the lowest spheroid mass of all ten ES
galaxies in the present sample. 
It is the only ES,e galaxy excluded due to its significant weight 
on the regression analysis of the ES,e/E galaxy types.  
NGC~3377 resides in 
the $M_{\rm bh}$-$M_{\rm *,sph}$ diagram (Figure~\ref{Fig1}) where one
would expect to find dust-poor S0 galaxies. 
Questions come to mind.  For example, 
might a large-scale disc be present, could the BH mass be in error,
could NGC~3377 have been built from a relatively rare collision between
disc-dominated galaxies in which the bulk of their net angular momentum
happened to cancel, or perhaps something else is afoot? 

Among the non-BCG E/ES galaxies, the E galaxy NGC~1600 is flagged but
retained.  It is a merger remnant and the dominant galaxy in a group
of 30 faint galaxies, which has been likened to a `fossil cluster'
\citep{2022MNRAS.509.2647R}.  
Given
its high BH and stellar mass, NGC~1600 likely formed from multiple major mergers
and perhaps should have been grouped with the BCGs.

NGC~6251 is the only E galaxy excluded from the sample
due to its outlying nature in the $M_{\rm bh}$-$M_{\rm *,sph}$ diagram.
Based on this, it may have formed from multiple galaxy mergers. 

NGC 3115 and NGC 6861 are dusty ES,b galaxies that have been included and
excluded from the regressions.  Having not grown a large-scale disc like the
dusty S0 galaxies, the ES,b galaxies \citep{Graham:Sahu:22b} reside to the
left of the $M_{\rm bh}$-$M_{\rm *,gal}$ relation defined by the non-BCG
merger-built sample.

\subsection{BCG}

Due to their low sample size of ten, a reliable linear regression was not
obtained for the BCG.  As such, none have been excluded {\it per se}, as none
were used in the present regressions.  However, it is noted that the Fornax
cluster BCG (NGC~1316) is a dusty S0 galaxy excluded from a previous
derivation of the BCG$+$(non-BCG) E$+$ES,e galaxy relation.  This exclusion
was because it was an S0 galaxy rather than an E or ES,e galaxy, but also due
to its outlying nature in the $M_{\rm bh}$-$M_{\rm *,gal}$ scaling diagram.
It appears to have been built by multiple major mergers.
With BH masses around $\sim$10$^9$ M$_\odot$,
the BCG NGC~7768 and the non-BCG NGC~6251 are two other candidates. 
Such galaxies open the door to three-body encounters, which may eject one of the
original BHs from the precursor galaxies \citep{2006ApJ...638L..75H}.
Of course, minor mergers may bring in a range of BHs. 
In passing, 
it is noted that while the Perseus cluster BCG NGC~1275 is an ES,e galaxy,
the other eight BCGs are E galaxies.

\bsp    
\label{lastpage}
\end{document}